\documentclass[a4paper,fleqn]{cas-dc}

\usepackage[authoryear]{natbib}
\usepackage{enumitem}
\usepackage{longtable}
\usepackage{booktabs}
\usepackage{graphicx}
\usepackage{subcaption}
\usepackage{caption}
\usepackage[most]{tcolorbox}
\usepackage{tabularx}
\usepackage{array}
\usepackage{listings}
\usepackage{xcolor}
\usepackage{tikz}

\colorlet{punct}{red!60!black}
\definecolor{background}{HTML}{EEEEEE}
\definecolor{delim}{RGB}{20,105,176}
\colorlet{numb}{magenta!60!black}

\definecolor{myblue}{RGB}{30,60,114}
\definecolor{lightblue}{RGB}{230,240,255}
\definecolor{lightgreen}{RGB}{240,255,235}
\definecolor{mygreen}{RGB}{60,100,70}
\definecolor{myorange}{RGB}{180,85,0}
\definecolor{lightorange}{RGB}{255,245,230}
\definecolor{mypurple}{RGB}{90,50,130}
\definecolor{lightpurple}{RGB}{245,235,255}

\tcbset{
  insightboxrq1/.style={
    enhanced,
    colback=lightgreen,
    colframe=lightgreen,
    boxrule=0pt,
    left=0pt,
    right=0pt,
    top=4pt,
    bottom=4pt,
    boxsep=5pt,
    borderline west={1pt}{0pt}{mygreen},
    arc=0mm,
    breakable
  },
  insightboxrq2/.style={
    enhanced,
    colback=lightorange,
    colframe=lightorange,
    boxrule=0pt,
    left=0pt,
    right=0pt,
    top=4pt,
    bottom=4pt,
    boxsep=5pt,
    borderline west={1pt}{0pt}{myorange},
    arc=0mm,
    breakable
  },
  insightboxrq3/.style={
    enhanced,
    colback=lightpurple,
    colframe=lightpurple,
    boxrule=0pt,
    left=0pt,
    right=0pt,
    top=4pt,
    bottom=4pt,
    boxsep=5pt,
    borderline west={1pt}{0pt}{mypurple},
    arc=0mm,
    breakable
  },
  insightboxrq4/.style={
    enhanced,
    colback=lightblue,
    colframe=lightblue,
    boxrule=0pt,
    left=0pt,
    right=0pt,
    top=4pt,
    bottom=4pt,
    boxsep=5pt,
    borderline west={1pt}{0pt}{myblue},
    arc=0mm,
    breakable
  }
}

\newcommand{\step}[1]{%
  \tikz[baseline=(char.base)]{%
    \node[shape=circle,fill=black,text=white,inner sep=1pt]
      (char) {\textbf{#1}};
  }%
}

\lstdefinelanguage{json}{
  basicstyle=\normalfont\ttfamily,
  numbers=left,
  numberstyle=\scriptsize,
  stepnumber=1,
  numbersep=8pt,
  showstringspaces=false,
  breaklines=true,
  frame=lines,
  backgroundcolor=\color{background},
  literate=
   *{0}{{{\color{numb}0}}}{1}
    {1}{{{\color{numb}1}}}{1}
    {2}{{{\color{numb}2}}}{1}
    {3}{{{\color{numb}3}}}{1}
    {4}{{{\color{numb}4}}}{1}
    {5}{{{\color{numb}5}}}{1}
    {6}{{{\color{numb}6}}}{1}
    {7}{{{\color{numb}7}}}{1}
    {8}{{{\color{numb}8}}}{1}
    {9}{{{\color{numb}9}}}{1}
    {:}{{{\color{punct}{:}}}}{1}
    {,}{{{\color{punct}{,}}}}{1}
    {\{}{{{\color{delim}{\{}}}}{1}
    {\}}{{{\color{delim}{\}}}}}{1}
    {[}{{{\color{delim}{[}}}}{1}
    {]}{{{\color{delim}{]}}}}{1}
}

\begin{document}

\shorttitle{Understanding Mobile App Recommendation Dynamics in General-Purpose LLMs}
\shortauthors{Motger et al.}

\title[mode=title]{Understanding Mobile App Recommendation Dynamics in General-Purpose LLMs: An Empirical Study}

\author[1]{Quim Motger}[orcid=0000-0002-4896-7515]
\cormark[1]
\ead{joaquim.motger@upc.edu}

\author[1]{Xavier Franch}[orcid=0000-0001-9733-8830]
\ead{xavier.franch@upc.edu}

\author[2]{Vincenzo Gervasi}[orcid=0000-0002-8567-9328]
\ead{vincenzo.gervasi@unipi.it}

\author[3]{Jordi Marco}[orcid=0000-0002-0078-7929]
\ead{jordi.marco@upc.edu}

\affiliation[1]{
  organization={Department of Service and Information System Engineering, Universitat Politècnica de Catalunya},
  city={Barcelona},
  country={Spain}
}

\affiliation[2]{
  organization={Department of Computer Science, University of Pisa},
  city={Pisa},
  country={Italy}
}

\affiliation[3]{
  organization={Department of Computer Science, Universitat Politècnica de Catalunya},
  city={Barcelona},
  country={Spain}
}

\cortext[1]{Corresponding author}

\begin{abstract}
Large Language Models (LLMs) are increasingly used to recommend mobile applications through natural language prompts, offering a flexible alternative to keyword-based app store search. 
Yet, the reasoning behind these recommendations remains opaque, raising questions about their consistency, explainability, and alignment with traditional App Store Optimization (ASO) metrics. 
In this paper, we present an empirical observational study of how general-purpose LLMs generate, justify, and rank mobile app recommendations across proprietary and open-source models, as well as knowledge-only settings with web search evaluated as a controlled ablation on the proprietary cohort. 
Our contributions are: (i) a taxonomy of 16 generalizable ranking criteria elicited from LLM outputs; (ii) a systematic evaluation framework to analyse recommendation consistency and the effect of explicit ranking instructions on cross-model convergence; and (iii) a replication package to support reproducibility and future research on LLM-based recommendation systems. 
Our findings reveal that LLMs report a broad yet fragmented set of ranking criteria, only partially aligned with standard ASO metrics. 
Proprietary models produce substantially more stable recommendations than locally deployed open-source models, and consistency varies substantially across app domains.
Furthermore, enabling web search does not materially change the ranking criteria reported by LLMs.
Contrary to our hypothesis, conditioning on explicit ranking criteria steers recommendations away from the blind baseline but reduces rather than increasing cross-model convergence. 
Our results aim to support end-users, app developers, and recommender-systems researchers in navigating the emerging landscape of conversational app discovery.\end{abstract}



\maketitle

\section{Introduction}
\label{sec:introduction}

Mobile apps have become central commodities in digital ecosystems, with billions of users relying on them for multiple features such as communication, productivity and entertainment~\citep{Statista2025}. Consequently, data-driven mobile app recommendation and ranking have received increasing attention in the field of software engineering, information systems and data mining, with research exploring multiple recommendation techniques such as collaborative filtering~\citep{Cao2017,He2017}, content-based filtering~\citep{Aliannejadi2021,Liptrot2024}, utility models~\citep{Dai2021}, machine learning models~\citep{Jisha2020} and, more recently, deep learning models~\citep{Aslam2022,Oyang2021} and neural architectures~\citep{Liu2023}. While technical advances in the context of mobile app recommender systems emerge, commercial app stores like Google Play\footnote{\url{https://play.google.com/store/apps}} and Apple's App Store\footnote{\url{https://www.apple.com/app-store/}} still lead the way as the largest global platforms for app distribution~\citep{Statista2025_numApps}.
Consequently, App Store Optimization (ASO) metrics (such as download counts, average ratings, user reviews, update frequency) have consolidated as key mechanisms to boost app visibility, perceived quality, and ranking~\citep{Lim2013,Picoto2019,Karagkiozidou2019,Papagiannis2020,Padilla-Piernas2020,Strzelecki2020}. These metrics simultaneously influence ranking algorithms and shape user perceptions of relevance, reliability, and overall quality~\citep{Lee01012014,XU2015171}.

Meanwhile, large language models (LLMs) have emerged as powerful interfaces for recommendation tasks~\citep{Zhao2024,Likang2024}, including software~\citep{Ma2024,John2024} and mobile app recommendation and discovery~\citep{Zhao2025}. 
LLMs can query, summarize, compare, and suggest mobile apps via natural language prompts, offering an alternative to keyword-based search and browsing in app stores. 
This shift reflects a broader trend in which users delegate exploratory search, comparison, and decision-making tasks to conversational agents, rather than interacting directly with platform-specific recommendation interfaces~\citep{Mo2025,TIBAU2024101331}.
However, the internal mechanisms for selection, ranking, and reporting used by LLMs remain opaque. 
As a result, their observable outputs constitute a novel and increasingly influential data source whose properties, biases, and consistency warrant systematic empirical analysis.
Beyond explainability, it remains uncertain whether their recommendations are grounded in empirical indicators of app quality, or instead reflect subjective, potentially non-reproducible and opaque reasoning.

Given the importance of transparent justifications and reproducibility in automated suggestions~\citep{Zhang2020}, this lack of transparency raises concerns. Users may rely on LLM-generated recommendations assuming they reflect real-world popularity, performance, or relevance. However, if such recommendations are inconsistent, non-evidence-based, or insensitive to context, they risk propagating misinformation and undermining trust. Likewise, developers and app providers may lack the knowledge or tools to improve their app's positioning in LLM recommendations, uncertain which aspects to prioritize, whether traditional positioning metrics still apply, and how to adapt to new LLM architectures and recommendation paradigms. While research on LLMs for query rewriting and content summarization has advanced~\citep{Ye2023}, little is known about how these models construct rankings for software artifacts such as mobile apps. In particular, the relationship between LLM-reported ranking criteria and established ASO signals remains unexplored.

In this paper, we empirically analyse through an observational study the logic, consistency, and transparency of mobile app recommendations generated by general-purpose LLMs. Our study covers ten models spanning two dimensions: {\it (i)} model openness, contrasting proprietary, API-based LLMs from leading commercial providers with open-source LLMs deployed locally on consumer-grade hardware; {\it (ii)}  information access, contrasting knowledge-only generation with a web-augmented setting evaluated as an ablation on the proprietary cohort.
First, we extract and structure app recommendations and the ranking criteria reported by LLMs when prompted for mobile app recommendations across feature-oriented searches (i.e., focused on a user-visible functional attribute of an app), drawn from two complementary feature datasets of different scale and category coverage~\citep{Dabrowski2023,Jiang2019}. 
Second, we assess the internal (per-model) and external (cross-model) consistency of the generated app recommendations. 
Third, we evaluate whether conditioning prompts on explicit ranking criteria steers models towards convergent recommendations, using the consistency of blind (i.e., unconditioned) recommendations as a baseline. 
Rather than evaluating recommendation correctness, our study focuses on inferring systematic patterns, assumptions, and decision signals embedded in LLM-generated outputs.
As a result, and to the best of our knowledge, we introduce the first systematic and replicable approach to study how general-purpose LLMs generate, justify, and rank mobile app recommendations across proprietary and open-source ecosystems and across web-augmented and knowledge-only settings. Specifically, our research provides the following contributions:

\begin{enumerate}[label=C$_{\arabic*}$]
    \item A taxonomy of ranking criteria explicitly reported by LLMs for mobile app recommendations, derived through prompt-based elicitation across diverse app domains, and validated across proprietary and open-source models, two feature datasets of different scale, and an ablation on web search access.
    \item A systematic empirical framework for observing, characterizing, and comparing LLM-generated recommendations, including analyses of internal and external consistency and an evaluation of whether explicit ranking criteria drive internal and cross-model convergence of recommendations.
    \item A replication package comprising code, prompts, datasets, and analysis scripts to facilitate reproducibility and extend the methodology and research framework to other software recommendation contexts, including local deployment configurations for open-source models.
\end{enumerate}

The rest of this paper is structured as follows. 
Section~\ref{sec:background} presents background literature and introduces terminology in the field of recommender systems, mobile app recommendation, ASO metrics and LLMs as recommender systems.
Section~\ref{sec:method} illustrates the research method, including research questions, use cases and the sample study methodology used in this research. 
Section~\ref{sec:elicitation} depicts the LLM ranking criteria elicitation process and results. 
Section~\ref{sec:consistency} reports the consistency analysis of mobile app recommendations. 
Section~\ref{sec:alignment} evaluates whether explicit ranking criteria steer LLMs towards convergent recommendations. 
Section~\ref{sec:discussion} synthesizes general findings and practical implications for users, developers and researchers. 
Section~\ref{sec:related-work} situates our study within existing literature on LLM-based recommendation. 
Finally, Section~\ref{sec:conclusions} summarizes contributions and outlines future research directions. 

All datasets, source code and evaluation reports are fully available\footnote{See Data Availability Statement at the end of this paper.}, enabling both replication studies and tracing the evolution of future LLMs as app recommenders.

\section{Background}
\label{sec:background}

\subsection{Mobile App Recommendation}

Recommender systems aim to match user needs with relevant items by modelling preferences, content, or behavioural patterns~\citep{Robillard2014Recommendation,Karumur2018}. Core approaches include collaborative filtering~\citep{Xue2019}, which infers preferences from peer behaviours; content-based filtering~\citep{Zhang2023}, which relies on item descriptors and user profiles; and hybrid methods~\citep{Ling2014} that combine both perspectives to improve coverage and mitigate cold-start issues, among others. These foundational models have evolved into more complex architectures, including (but not limited to) matrix factorization~\citep{Kim2016}, neural networks~\citep{Wu2022} and, more recently, predictive LLMs like BERT~\citep{Yang2022} and generative LLMs like GPT-4~\citep{Zhao2024,TIBAU2024101331}, offering varying degrees of scalability, personalization, and explainability. These generative LLMs increasingly serve as recommendation and discovery interfaces~\citep{Mo2025}, either explicitly, through direct conversational interaction, or implicitly, embedded behind traditional search interfaces to synthesize and re-rank results without a user-facing conversational exchange, as in Google's AI Mode\footnote{\url{https://search.google/ways-to-search/ai-mode/}}.

Mobile app recommendation builds on these methods but introduces domain-specific challenges. Unlike static items such as movies or books, mobile apps are frequently updated, have variable functionality across devices, and are typically searched with concrete goals in mind, such as completing a task or solving a problem. User expectations are highly contextual (e.g., ``\textit{a lightweight photo editor with offline mode}''), and so recommendation models must account for metadata, technical specifications, user feedback, and temporal dynamics~\citep{Liu2016,He2017,Ferretto2017,Bendada2020}. This has led to the need for mobile app recommendation pipelines that incorporate structured descriptors (e.g., permissions, categories)~\citep{Zhu2014}, unstructured content (e.g., user reviews, changelogs)~\citep{Lin2018}, and behavioural data (e.g., evolution of downloads)~\citep{Zhong2019}.

At the same time, commercial app stores such as Google Play and the Apple's App Store offer their own ranking and recommendation systems\footnote{\url{https://support.google.com/googleplay/android-developer/answer/9958766?hl=en}, \url{https://developer.apple.com/app-store/search/}}, designed to surface apps that align with user interests, trending behaviour, and strategic promotion. These systems are opaque by design, with ranking logic influenced by proprietary algorithms, monetization strategies, and curated features such as top charts and editor's choices~\citep{Farooqi2020,Bello2020}. Prior studies have attempted to audit these stores via crawls and ranking comparison~\citep{Lim2013,Carbunar2015,Wu2018,Wang2018,Comino2019Updates,Li2024}, but these efforts face inherent replicability challenges due to limited transparency and volatile interfaces. As a result, much of the existing work in this area relies on observational analyses of system outputs rather than direct access to underlying ranking mechanisms.

These studies have motivated interest in evaluating not only which apps are recommended, but also why certain rankings are produced, under what assumptions, and whether the underlying logic can be traced, reproduced, or contested (see Section~\ref{sec:aso}). On the other hand, the rise of generative LLMs introduces new recommendation scenarios, as well as methodological and interpretation challenges beyond traditional techniques (see Section~\ref{sec:llm-rec}).

\subsection{App Store Optimization (ASO)}
\label{sec:aso}

App Store Optimization (ASO) refers to the practice of improving an app’s discoverability and ranking within app store ecosystems by optimizing key performance and metadata signals~\citep{Lee01012014,Karagkiozidou2019,ABRAR2025100292}. Common ASO metrics include average user ratings~\citep{Shen2017}, total downloads~\citep{Chen2017}, number and sentiment of user reviews~\citep{Karagkiozidou2019}, update frequency~\citep{Lim2013}, keyword relevance~\citep{Li2024}, and language availability~\citep{Picoto2019}. These metrics influence both user visibility and backend ranking algorithms, although the exact weight and interaction of these factors remain undocumented and may vary across app categories, search terms, localizations, user profiles and platforms~\citep{Lim2013,Carbunar2015}. For developers, ASO serves as a tactical lever to improve market performance and user acquisition, often aligned with marketing and release strategies. For users, these metrics act as interpretive signals, heuristics that influence trust, expectations, and app selection. Several empirical studies have explored how users perceive ASO elements~\citep{Strzelecki2020,Alhejaili2022}, how ASO metrics correlate with app success~\citep{Karagkiozidou2019}, and how the manipulation of these signals (e.g., via fake reviews, download farms) distorts platform dynamics~\citep{Chen2017,Martens2019}.

In addition to behavioural phenomenon analysis, ASO has been also studied as a prediction target for app success and ranking. Models have been proposed to forecast app popularity from early ASO signals~\citep{ABRAR2025100292}, detect fake reviews or rating inflation~\citep{Hernandez2021}, and assess the fairness and transparency of ranking mechanisms~\citep{Brouwer2020}. However, these efforts typically focus on app stores as closed ecosystems. Conversational LLMs, by contrast, may synthesize recommendations without direct access to or explicit use of these structured signals, instead relying on inferred, implicit, or qualitative cues. To the best of our knowledge, there are no published studies on how traditional ASO metrics might be interpreted --- or entirely bypassed --- by new interfaces like conversational LLMs. When LLMs recommend apps without referencing these signals, or cite alternative criteria altogether, it becomes unclear whether conventional metrics retain relevance, or whether new interpretive frameworks are needed to understand ranking behaviour.

\subsection{LLMs as Recommender Systems}
\label{sec:llm-rec}

LLMs have rapidly evolved from general-purpose language tools into powerful agents capable of answering domain-specific questions, performing structured reasoning, and generating informative responses based on content retrieval~\citep{Raiaan2024}. Their integration into recommendation workflows has enabled zero-shot conversational interactions that leverage both content and contextual signals to generate recommendations~\citep{Jannach2021}. Recent work on conversational search shows that users increasingly rely on dialogue-based interactions with LLMs to explore alternatives, refine criteria, and support decision-making, rather than issuing isolated keyword queries~\citep{Mo2025,TIBAU2024101331}. Unlike traditional recommender systems, relying on predefined databases, user histories, and model tuning, LLMs operate in a generative and context-dependent approach, producing both ranked outputs and natural language justifications on demand~\citep{He2023}. However, such justifications may reflect post-hoc rationalizations rather than faithful representations of the internal decision process\citep{Turpin2023,Lanham2023}, complicating interpretation and trust. 

In the mobile app domain, LLMs can generate recommendation lists for queries like ``\textit{What's a secure messaging app for Android?}'' or ``\textit{Recommend an app to listen to podcasts}'' by synthesizing app names, features, comparative statements, and quality indicators. To satisfy these queries, LLMs may operate under two distinct information access paradigms. 
In a \textit{knowledge-only} setting, recommendations derive exclusively from parametric knowledge internalized at training time, making outputs self-contained but potentially outdated or ungrounded~\citep{Raiaan2024}. 
In a \textit{web-augmented} setting, retrieval-enhanced services such as ChatGPT\footnote{\url{https://openai.com/index/introducing-chatgpt-search/}} and Gemini\footnote{\url{https://ai.google.dev/gemini-api/docs/google-search}} combine internal knowledge with live web data through search-based pipelines, creating the perception of interaction with an up-to-date domain expert~\citep{DiPalma2023}, at the cost of introducing undisclosed, volatile retrieval steps that further complicate transparency and reproducibility, as well as additional computational and economic overhead per query.
Orthogonally, LLMs differ in openness: proprietary models are accessible only through hosted APIs, whereas open-source models can be deployed locally under fully controlled, replicable conditions.

Under either paradigm, general-purpose LLMs can serve as stand-alone recommender engines when guided by carefully designed prompts and strategies.
Xu et al.~\citep{Xu2025} proposed a comprehensive framework for employing LLMs as recommender systems, outlining key components such as prompt engineering, tuning strategies, and candidate item (e.g., suggested app) construction. Several challenges identified in their study are directly relevant to the mobile app domain. 
Recommendations may include hallucinated or obsolete app names; furthermore, the inherent variability of LLM outputs due to model architecture, prompt phrasing, or context length complicates reproducibility and interpretability. 
Xu et al.~\citep{Xu2025} further highlight the difficulty LLMs face in retaining structured item lists, as well as their tendency to deviate from intended recommendation goals. 
While prompting strategies such as few-shot examples and role-based instructions offer some control, they are insufficient for ensuring alignment with real-world quality indicators, especially for subjective, qualitative characteristics such as the app's ``ease of use'' or ``security'', lacking evidence for grounding or standardization. These limitations underscore the need for systematic, observational empirical evaluations of LLM-based app recommendations, particularly regarding their stability, consistency, and the relationship between reported criteria and observable recommendation behaviour.
\section{Method}
\label{sec:method}

\subsection{Goal and Research Questions}

The goal of this research is \textbf{to examine how general-purpose conversational LLMs generate mobile app recommendations, focusing on the reasoning behind their outputs}. 
More broadly, we aim to support transparent, consistent, and evidence-based use of LLMs in software recommendation tasks. 
While the methodology is applicable to a broad range of LLM-based recommenders, our study spans two dimensions of the current LLM landscape: model openness, covering both proprietary, API-based LLMs and open-source LLMs deployed locally on consumer-grade hardware; and information access, covering knowledge-only setting relying exclusively on parametric knowledge as the main cohort as well as web-augmented generation as an ablation study to assess its impact on recommendation consistency and sensitiveness to ranking criteria. 
To this end, we design our study as an empirical observational study using real-world data within the context of mobile apps across multiple categories and domains (see Section~\ref{sec:sample-study}).

Specifically, we seek to answer the following research questions (RQ):

\begin{enumerate}[label=\textbf{RQ$_{\mathbf{\arabic*}}$}]
    \item Which ranking criteria do LLMs claim to use when recommending mobile apps?
    \item How consistent are LLM mobile app recommendations, within and across models?
    \item To what extent does conditioning on explicit ranking criteria steer LLMs towards convergent recommendations?
\end{enumerate}



To answer \textbf{RQ\textsubscript{1}}, we designed a systematic prompt-based multi-model experiment to extract mobile app recommendations and elicit the factors that influence their ranking. 
First, we conducted a literature analysis to identify standard ASO metrics, which served as grounding for analysing the ranking criteria used by LLMs. 
Then, following established guidelines for prompt-based recommender systems~\citep{Zhao2024,Xu2025}, each model was prompted with feature-based searches to recommend a fixed number of popular apps. 
The outputs were parsed and post-processed to extract app recommendations and ranking criteria. 
A bottom-up semi-automatic approach was used to synthesize these reported rationales into a taxonomy of generic ranking criteria articulated by LLMs for the mobile app domain, which we finally aligned with ASO metrics to support a comparative analysis. 
We conducted this elicitation under four experimental configurations, crossing the two model cohorts (proprietary, open-source) with the two feature datasets (small, large), and consolidated the reported rationales into a single taxonomy while reporting convergence across configurations. Additionally, we ran an ablation on information access, replicating the elicitation with web search enabled for the proprietary cohort on the small-scale dataset, to assess whether web search significantly modifies the elicited ranking criteria (see Section~\ref{sec:information-access}).

To answer \textbf{RQ\textsubscript{2}}, we conducted a consistency analysis of LLM app recommendations along two dimensions: internal consistency (i.e., variability across multiple runs of the same LLM) and external consistency (i.e., variability across different LLMs). 
To stress consistency under a broad and heterogeneous set of recommendation contexts, this analysis is conducted on the large-scale feature dataset; the same consistency metrics are later reused in RQ\textsubscript{3}, where blind recommendations (i.e., generated without specifying a ranking criterion) on the small-scale dataset serve as the convergence baseline.
As in RQ\textsubscript{1}, we complement this analysis with an information access ablation, comparing consistency scores with and without web search for the proprietary cohort on the small-scale dataset.

To answer \textbf{RQ\textsubscript{3}}, we conducted an extended systematic prompt-based experiment by explicitly feeding specific ranking criteria to the LLMs. 
Each ranking criterion from the taxonomy derived in RQ\textsubscript{1} was used as a search condition in the user prompt. Our working hypothesis is that explicit ranking criteria act as a shared reference that reduces model-specific interpretation, steering different LLMs towards convergent recommendations. 
To test this, we use the consistency of blind recommendations as a baseline, and measure whether guided recommendations (i.e., generated with a specific ranking criterion) exhibit higher cross-model agreement. 
Since RQ\textsubscript{3} measures the relative change in cross-model agreement between guided and blind recommendations over the same feature set, its validity depends on the heterogeneity of the queried domains rather than on the number of features; the small-scale dataset, covering 8 distinct app categories, satisfies this requirement while keeping the feature $\times$ criteria $\times$ runs $\times$ models cross-product tractable, particularly for locally deployed models (see Section~\ref{sec:sample-study}).

\subsection{Use Cases}

We highlight three representative use cases illustrating how different stakeholders can benefit from the results of this study. These use cases are referenced throughout the paper to contextualize the discussion of the research questions, findings, and practical implications.

\begin{itemize}

    \item \textbf{UC1: Informed app choice through transparent LLM recommendations} (\textit{Stakeholders: End-users}). As users increasingly rely on LLMs to discover mobile apps, the lack of clarity behind recommendations may affect trust and decision-making. Our findings help users interpret the rationale behind LLM recommendations, assess consistency within and across LLM providers, and understand how model openness and web search access shape recommendation behaviour.

    \item \textbf{UC2: Optimizing app visibility in LLM-driven ecosystems} (\textit{Stakeholders: App developers and ASO specialists}). LLMs are emerging as influential platforms for app discovery, complementing traditional app stores. By revealing the ranking criteria that LLMs apply and their relative influence, this study equips developers with actionable knowledge to optimize app presentation, metadata, and positioning for improved discoverability in LLM-generated lists.

    \item \textbf{UC3: Designing transparent and controllable app recommender systems.} (\textit{Stakeholders: Recommender system researchers}). LLM-based recommendations pose new challenges for consistency, alignment, and justification. Our study contributes empirical evidence and methodological tools to support the design of LLM-based app recommenders that are more interpretable, verifiable, and responsive to user intents.

\end{itemize}

\subsection{Sample Study: Feature-based Search with General-Purpose LLMs}
\label{sec:sample-study}  

We shaped this research as a \textit{sample study}, following the strategy framework proposed by Stol and Fitzgerald~\citep{Stol2018}. 
Our goal is to maximize potential for \textit{generalizability} over actors (i.e., LLM providers and models), following an observational sampling perspective. 
Consequently, we focused on minimizing \textit{obtrusiveness} (i.e., intervention on real world data and experiment settings) by leveraging existing search datasets on popular mobile apps and domains, as well as established, simple prompt patterns for recommender systems. 
On the other hand, we pursue generalizability by focusing on universal contexts and systems: we sample both proprietary LLMs from leading commercial providers and open-source LLMs from major model families, and we rely on the default configuration of each LLM, reflecting the typical user experience when interacting with these systems in real-world scenarios.

Our study deliberately does not adopt a recommendation baseline. We observe and characterize the behaviour of widely adopted, general-purpose LLMs under comparable conditions, treating each model as an independent observational unit; introducing a baseline would impose an artificial reference and introduce selection bias, which is incompatible with the observational nature of the study. Furthermore, we do not compare LLM-based recommendations with classical recommender systems or app store rankings, as our focus is on analysing LLM-specific properties such as recommendation consistency, cross-model agreement, variability across runs, and sensitivity to ranking criteria, rather than relative effectiveness.

Below we present the specifics of this research design into the three key variables of our sample study: (1) datasets, (2) model selection, and (3) information access settings.

\subsubsection{Datasets}

\paragraph{Small-scale dataset.} We used a subset of the dataset from D\k{a}browski et al.~\citep{Dabrowski2023}, comprising 124 app features elicited by human annotators from the descriptions of 8 popular mobile apps, each belonging to a distinct Google Play category. From these, we selected two features per category through manual inspection, excluding overly generic (e.g., \textit{sync}), app-specific (e.g., \textit{use netflix}), or ambiguous (e.g., \textit{widget}) items, and randomly sampling among the remaining candidates. This yielded 16 features from 8 domains, each treated as a representative proxy for a broader functional domain. This set is used for the ranking criteria elicitation and its web-search ablation (RQ\textsubscript{1}), and for the criteria-conditioned experiments (RQ\textsubscript{3}), where the combinatorial cost of crossing features, criteria, runs, and models constrains the feasible dataset size.

\paragraph{Large-scale dataset.} To assess whether our findings generalize beyond the small-scale dataset, 
we additionally sampled the dataset by Jiang et al.~\citep{Jiang2019}, comprising 533 app features elicited from mobile app descriptions across 18 Google Play categories. We applied stratified random sampling over app categories to select 100 features, preserving the category distribution of the original dataset while keeping the experimental cost of repeated multi-model querying tractable. This dataset is used to replicate the ranking criteria elicitation (RQ\textsubscript{1}) and as the primary dataset for the consistency analysis (RQ\textsubscript{2}).

\subsubsection{Model selection}

We selected ten LLMs organized in two cohorts: proprietary models accessed via API, and open-source models deployed locally (see Table~\ref{tab:sample-models}).

For the proprietary cohort, we use popularity, accessibility for research, and relevance to the scientific community as selection criteria. With respect to popularity, we consider up-to-date statistical reports on AI chatbot market shares worldwide~\citep{statcounter2025chatbots}. Concerning accessibility, we require: (i) API access for large-scale experimentation; (ii) support for structured (i.e., schema-constrained) outputs, enabling reliable parsing of recommendations and criteria at scale; and (iii) availability of web search as an API-level tool, required for the information access ablation in RQ\textsubscript{1} and RQ\textsubscript{2}  (see Section~\ref{sec:information-access}). These criteria led to the exclusion of some providers: Perplexity's Sonar failed to reliably adhere to structured output constraints in preliminary tests, producing frequent schema violations (e.g., inconsistent list sizes), while DeepSeek did not offer web search functionality through its API at the time of the experiments. For each selected provider (OpenAI, Google, Anthropic, and Mistral), we selected the flagship conversational model available at the time of the experiments.

For the open-source cohort, 
the selection followed three criteria: (1) one instruction-tuned chat model per major open-source family; and (2) the highest-ranked variant on the Artificial Analysis Intelligence Index as of May 27th, 2026 that could be executed on our hardware. This resulted in six models from the Llama, Gemma, Qwen, GPT-OSS, Mistral, and DeepSeek families. All models were served locally through Ollama\footnote{\url{https://ollama.com/}} on a workstation equipped with an NVIDIA GeForce RTX 4090 GPU (24~GB VRAM), an AMD Ryzen~9 7950X CPU (16 cores), and 64~GB of RAM, running Ubuntu 24.04 LTS; the 24~GB VRAM budget acts as the hardware constraint in criterion (2), bounding the maximum model size per family.

Across both cohorts, we kept all generation settings at each model's default values, reflecting the typical user experience when interacting with these systems.

\begin{table}[t]
\centering
\caption{Sample study: selected models per cohort}
\label{tab:sample-models}
\begin{tabular}{lll}
\toprule
\textbf{Cohort} & \textbf{Provider} & \textbf{Model} \\
\midrule
Proprietary & OpenAI & gpt-5.3-chat \\
Proprietary & Google & gemini-3-flash \\
Proprietary & Anthropic & claude-opus-4-6 \\
Proprietary & Mistral & mistral-large-2512 \\
\midrule
Open-source & Meta & llama4:scout \\
Open-source & Google & gemma3:27b \\
Open-source & Alibaba & qwen3:30b-a3b \\
Open-source & OpenAI & gpt-oss:20b \\
Open-source & Mistral & mistral-small3.1:24b \\
Open-source & DeepSeek & deepseek-r1:8b \\
\bottomrule
\end{tabular}
\end{table}

\subsubsection{Information access settings}
\label{sec:information-access}

We consider two information access settings: a \textit{knowl\-edge-only} setting, in which models generate recommendations exclusively from parametric knowledge, and a \textit{web-augmented} setting, in which models are additionally granted access to their provider's native web search tool via API. 
The web-augmented setting is available only for the proprietary cohort, as the selected open-source models do not natively integrate web search and equipping them with external retrieval pipelines would introduce uncontrolled, non-comparable tooling.

We treat information access as an ablation variable: proprietary models are run under both settings on the small-scale dataset, allowing us to assess the effect of web search on the two objects of our analysis — the elicited ranking criteria (RQ\textsubscript{1}) and the consistency of app recommendations (RQ\textsubscript{2}). 
Since our study does not assess the correctness of recommendations nor the identity of candidate apps, these two ablations suffice to determine whether web search materially affects our scope of analysis. Based on their outcome, and given the substantial time, energy, and economic cost of web-augmented querying at scale, all remaining experiments (RQ\textsubscript{3}) are conducted in the knowledge-only setting for all models, in line with responsible and resource-efficient experimentation.
\section{Elicitation of Ranking Criteria (RQ\textsubscript{1})}
\label{sec:elicitation}

\subsection{Design}

The elicitation and categorization of LLM ranking criteria is structured in two main processes: the elicitation of ASO metrics from literature (steps \step{1} and \step{2}), and the elicitation of LLM ranking criteria from a systematic prompt-based experimentation (steps \step{3} to \step{6}). 
Results from both processes are reported in Section~\ref{sec:rq1-results}.


\subsubsection{ASO Metrics}

We structured the extraction of ASO metrics from background literature in two main stages:

\begin{itemize}
    \item \textbf{Step \step{1} - Literature review.} We focused on literature in the ASO field and related terms to identify peer-reviewed publications defining ASO metric taxonomies. Table~\ref{tab:literature-review} includes the details of the search strategy. For the search string, we piloted query variants combining \textit{app store optimization} and related terms used in the literature (see Section~\ref{sec:background}). For databases, we focused on peer-reviewed publications from publishers within the computer science field, excluding pre-prints and non-peer-reviewed publications. We used our search string to retrieve studies from each of these publishers. After study collection, we applied the following steps: (1) study deduplication, (2) assessment of inclusion/exclusion criteria, and (3) study selection. We did not apply snowballing as the first selection of studies already included a systematic literature review that applied snowballing and covered foundational and related works on ASO metrics~\citep{Karagkiozidou2019}.

    \item \textbf{Step \step{2} - ASO metrics extraction.} We extracted and categorized ASO metrics from the selected studies, resulting in a taxonomy that includes, for each node (metric): metric’s unique identifier, name, description, data type, and origin (i.e., whether user-, developer-, or third-party-dependent).
\end{itemize}

\begin{table}[t]
\centering
\caption{Search strategy}
\label{tab:literature-review}
\renewcommand{\arraystretch}{1.3}
\begin{tabularx}{\columnwidth}{>{\raggedright\arraybackslash}p{1.2cm} X}
\toprule
\textbf{Search String} &
\texttt{("app store optimization" OR "app store ranking" OR "app ranking optimization" OR "app visibility optimization" OR ("ASO" AND ("app store*" OR "mobile app*" OR "app visibility*")))} \\
\midrule
\textbf{Databases} &
Scopus, IEEE Xplore, ACM Digital Library, Springer Link, Science Direct \\
\midrule
\textbf{Inclusion Criteria} &
(1) The paper explicitly discusses app performance/visibility/ranking optimization \newline (2) The paper defines quantitative or qualitative metrics related to app optimization \\
\midrule
\textbf{Exclusion Criteria} &
(1) The paper is out of scope \newline (2) Full text not available \newline (3) Full text not in English \\
\bottomrule
\end{tabularx}
\end{table}

\subsubsection{LLM Ranking Criteria}

Based on the framework by Xu et al.~\citep{Xu2025}, we designed our approach as a \textit{prompting without tuning} strategy (i.e., default configuration from general-purpose generative LLMs, both proprietary and open-source) on a \textit{zero-shot ranking task} (i.e., mobile app recommendation without examples or historical records). Our analysis focused on a \textit{beyond-accuracy} recommendation task, aiming at providing \textit{explainable} recommendations through the elicitation of ranking criteria used to provide such recommendations. We used related work covered by Xu et al.~\citep{Xu2025} and other systematic literature reviews in the field~\citep{Zhao2024,Likang2024,Zhao2025} to design \textit{system} and \textit{user} prompts used in our experiments. For the system prompt, we used a  \textit{role prompting} strategy applying the persona pattern for mobile app market experts.

\begin{tcolorbox}[
colback=gray!5!white,
colbacktitle=black,
coltitle=white,
title=\textbf{System prompt},
boxrule=0pt,
left=0.5mm, right=0.5mm, top=0.5mm, bottom=0.5mm,
sharp corners,
enhanced
]
You are a mobile app market expert specializing in analysing and ranking apps. Your task is to identify and rank a given number of apps for a particular feature based on your own ranking criteria. A ranking criterion refers to a specific, measurable, and traceable metric or characteristic of an app that highly impacts its position in your ranking.
\end{tcolorbox}

For the user prompt\footnote{The example in grey text is not included in the actual prompt; it is shown in the manuscript for explanatory purposes.}, we use a top-$k$ recommendation template with a search-based \texttt{feature} extension~\citep{Xu2025}.

\begin{tcolorbox}[
colback=gray!5!white,
colbacktitle=green!50!black,
coltitle=white,
title=\textbf{User prompt (RQ\textsubscript{1})},
boxrule=0pt,
left=0.5mm, right=0.5mm, top=0.5mm, bottom=0.5mm,
sharp corners,
enhanced
]
Recommend \texttt{\{k\}} apps to \texttt{\{feature\}}. \textcolor{gray}{\quad\#\quad E.g., Recommend 20 apps to build photo collages.}
\end{tcolorbox}

Whenever possible, we configured the LLM via API parameter settings or, failing that, used an extended system prompt to ensure responses consistently conformed to the following JSON schema:

\begin{lstlisting}[language=json,
  firstnumber=1,
  basicstyle=\ttfamily,
  columns=fullflexible,
  literate={//}{{\ //\ }}2,
  xleftmargin=0.5em,
  xrightmargin=0.5em,
  aboveskip=2pt,
  belowskip=2pt,
  lineskip=-1pt]
{
  "a": [ // App names ordered by rank
    "App A",
    "App B",
    "App C" 
  ],
  "c": [ // Additional ranking criteria
    {
      "n": "Name of the ranking criterion",
      "d": "Description of the ranking criterion"
    } 
  ]
}
\end{lstlisting}

We used the feature dataset and the selected LLMs (from Section~\ref{sec:sample-study}) as well as the previous prompts as input to conduct our systematic prompt-based analysis to extract app recommendations and ranking criteria. We designed this approach as a semi-automatic process, with minimum human intervention, agnostic to the mobile app domain (with the exception of the prompts and dataset), to facilitate both the replication of our study and the generalization to other domains. This process is structured as follows:

\textbf{Step \step{3} - LLM criteria extraction}. We developed a multi-agent software service to systematically prompt multiple LLM providers and models, both via API (proprietary cohort) and via local deployment (open-source cohort). We set $k=20$, a common setting in recommender system evaluation~\citep{Hidasi2018,Chen2022}. 
Each \texttt{feature} was prompted across 10 independent runs to account for variability and consistency of app recommendations (see Section~\ref{sec:consistency}). We executed this extraction under five experimental configurations (see Section~\ref{sec:sample-study}): the four combinations of model cohort (proprietary, open-source) and feature dataset (small, large) in the knowledge-only setting, plus the web-augmented configuration for the proprietary cohort on the small-scale dataset, used for the information access ablation. 
We then parsed and consolidated all app recommendations and ranking criteria across all runs and configurations.

\textbf{Step \step{4} - LLM criteria filtering}. We applied three automatic, conservative filtering steps to reduce noise and redundancy, avoiding excessive manipulation and preserving generalization of our study. First, we performed exact-match deduplication based on both the name and description of each ranking criterion, removing only fully identical entries. Second, we removed criteria whose name appeared only once across all runs (i.e., across all models and features), assuming that such isolated names likely reflect uncommon or non-representative outputs. Third, we removed criteria whose name appears exclusively within a single feature, thus excluding overly narrow or domain-specific naming patterns that do not generalize beyond one context.

\textbf{Step \step{5} - LLM criteria consolidation}. After filtering, we applied a hierarchical clustering process to simplify the consolidation and deduplication process of non-exact matches, given the large variability in phrasing and even naming of the ranking criteria. This process was structured as follows:
\begin{enumerate}
    \item We used a Sentence-BERT~\citep{Reimers2019} model (\texttt{all-MiniLM-L6-v2}) to generate sentence embeddings from each name-description ranking criteria pair. We selected this BERT modification due to its proved performance for deduplication tasks in the context of short sentences~\citep{Isotani2021}, both in terms of efficiency and accuracy. 
    \item We normalized the embeddings and computed the cosine similarity between all ranking criteria pairs to create a similarity matrix. 
    \item To determine the optimal number of clusters, we implemented a systematic multi-faceted clustering approach, as also applied in other domains~\citep{Jin2025}, using four complementary methods: silhouette analysis~\citep{Dinh2019}, gap statistic~\citep{Tibshirani2002}, elbow method~\citep{Shi2021}, and Calinski-Harabasz index~\citep{Caliński01011974}. These scores were normalized, aggregated and equally balanced to select the optimal cluster count. We then performed hierarchical clustering using Ward linkage, which minimizes within-cluster variance~\citep{Strauss2017}. 
    \item In order to select a representative ranking criterion from each cluster, we employed a voting mechanism combining three centrality measures: (1) centroid distance (minimum Euclidean distance to cluster centroid), (2) connectivity score (average cosine similarity to all cluster members), and (3) silhouette contribution (highest individual silhouette score). The most frequently selected candidate across these three methods was chosen as the cluster representative. 
    \item After automatic analysis, we manually inspected the resulting list to (1) exclude domain-specific criteria tied to a particular app category (e.g., \textit{content library size} for streaming apps), and (2) deduplicate any remaining redundant entries by selecting through agreement a representative for each set of duplicated ranking criteria.
\end{enumerate}

\textbf{Step \step{6} - Cross-configuration analysis}. The consolidation pipeline (Steps \step{4}--\step{5}) is applied to the pooled knowledge-only outputs of all four cohort--dataset configurations, yielding a single consolidated taxonomy. We then assess convergence by measuring the coverage of each consolidated criterion across configurations (i.e., whether each criterion is reported by both cohorts and in both datasets). Finally, for the information access ablation, we apply the same pipeline to the web-augmented outputs of the proprietary cohort on the small-scale dataset and compare the resulting criteria against their knowledge-only counterparts, assessing whether web search materially modifies the elicited ranking criteria.

\subsection{Results}
\label{sec:rq1-results}

For the extraction of ASO metrics (Steps \step{1} $\rightarrow$ \step{2}), we collected 28 studies using the search string and databases identified in Table~\ref{tab:literature-review}. We removed 8 duplicated studies and 14 additional studies which either did not match both inclusion criteria or matched one of the exclusion criteria. This led to a total selection of six studies (including one systematic literature review~\citep{Karagkiozidou2019}) defining ASO metric taxonomies, with different degrees of extension and formalization. We inspected these studies and consolidated a unified taxonomy of 23 distinct ASO metrics, which we report in Table~\ref{tab:aso-metrics}.

\begin{table}[t]
\centering
\scriptsize
\setlength{\tabcolsep}{3pt}
\caption{Description of App Store Optimization (ASO) metrics}
\label{tab:aso-metrics}
\begin{tabularx}{\columnwidth}{@{}lp{1.5cm}X@{}}
\toprule
\textbf{ID} & \textbf{Name} & \textbf{Description} \\
\midrule
ASO\_01 & Icon & App icon. \\
ASO\_02 & Media & App images/videos in app store description. \\
ASO\_03 & Description & Written description of the app. \\
ASO\_04 & Rating & Average rating from users in user reviews. \\
ASO\_05 & Reviews & Total number of user reviews. \\
ASO\_06 & Rank & Position on top chart's app category. \\
ASO\_07 & Downloads & Number of app downloads. \\
ASO\_08 & Downloads per Day & Number of app downloads (previous day). \\
ASO\_09 & Reviews Content & Positive/negative impressions from users. \\
ASO\_10 & Localization & Languages/localizations available on the app. \\
ASO\_11 & Name & App name/title. \\
ASO\_12 & Genre & App genre. \\
ASO\_13 & URL & Available external URL. \\
ASO\_14 & Last Update Date & Date of last update. \\
ASO\_15 & Update List & List of updates. \\
ASO\_16 & System Version & Operating system version requirement. \\
ASO\_17 & Developer & Name of the developer. \\
ASO\_18 & Content Rating & Age group suitable to use the mobile app. \\
ASO\_19 & Changelog & Description of changes in last update. \\
ASO\_20 & Price & Price to purchase or access app. \\
ASO\_21 & Category & App category. \\
ASO\_22 & Release Date & The release date of the app. \\
ASO\_23 & Rating Distribution & The number of \{1,2,3,4,5\} ratings of the app. \\
\bottomrule
\end{tabularx}
\end{table}

To elicit LLM ranking criteria (Steps \step{3} $\rightarrow$ \step{5}), the systematic prompt-based analysis (Step \step{3}) across the four knowledge-only cohort--dataset configurations, produced 76,119 raw ranking criteria items from 11,600 LLM queries. 
The filtering stage (Step \step{4}) removed redundancies and infrequent items through three main operations: (i) exact name-description duplicates (e.g., \emph{Ease of Use - How intuitive and streamlined the user interface is}), with 3,352 items removed; (ii) exclusion of unique criteria names cited only once across all queries (e.g., \emph{Delivery Reliability and Speed - The consistency and speed with which broadcast messages are successfully delivered}), with 13,284 removed; and (iii) exclusion of rare, feature-specific criteria with unique names within individual app features (e.g., \emph{Broadcast Reach Capacity - The maximum number of recipients an app can support in a single broadcast}), with 9,291 removed. 
This conservative filtering reduced the set to 50,192 criteria. 
In the consolidation stage (Step \step{5}), hierarchical clustering was applied per configuration, producing 180 cluster representatives (50, 47, 40, and 43 for the proprietary--small, proprietary--large, open--small, and open--large configurations, respectively). 
Manual review excluded domain-specific representatives (e.g., \emph{Content Catalog Depth}, \emph{Streaming Quality}), yielding 54 configuration-level criteria, which we consolidated through cross-configuration semantic alignment (e.g., merging \emph{Cross-platform Compatibility - Number of supported platforms and operating systems} with \emph{Platform Availability - The operating systems and devices the app supports}) into a final unified taxonomy of 16 distinct and generalizable LLM ranking criteria. 
Table~\ref{tab:llm-aso-mapping} shows these criteria and their alignment with ASO metrics where applicable. 

\begin{table}[t]
\centering
\scriptsize
\setlength{\tabcolsep}{3pt}
\caption{Description of LLM ranking criteria and alignment with ASO metrics}
\label{tab:llm-aso-mapping}
\begin{tabularx}{\columnwidth}{@{}lp{1.5cm}Xl@{}}
\toprule
\textbf{ID} & \textbf{Name} & \textbf{Description} & \textbf{ASO} \\
\midrule
LLM\_01 & Accessibility Features & Accessibility support for users with disabilities (e.g., screen readers, captions, inclusive design). & -- \\
LLM\_02 & Cost and Pricing & Pricing model, including free tiers, one-time purchases, subscriptions, in-app purchases, and value for money. & ASO\_20 \\
LLM\_03 & Cross-Platform Synchronization & Consistency and synchronization of data and experience across devices and platforms. & -- \\
LLM\_04 & Customer Support & Availability, responsiveness, and quality of customer support. & -- \\
LLM\_05 & Customization Options & Extent to which users can personalize appearance, themes, layout, and behaviour. & -- \\
LLM\_06 & Feature Breadth and Richness & Range, depth, and quality of features, including core and optional capabilities. & -- \\
LLM\_07 & Integration with Other Services & Compatibility and integration with other apps and platforms (e.g., email, cloud storage, social media). & -- \\
LLM\_08 & Performance and Stability & Speed, responsiveness, and reliability, including stable operation without crashes or lag. & -- \\
LLM\_09 & Platform and Device Compatibility & Support across operating systems, devices, and platforms, affecting accessibility. & ASO\_16 \\
LLM\_10 & Popularity and Market Reach & Adoption and audience size, including downloads and active users. & ASO\_07 \\
LLM\_11 & Regional and Geographical Availability & Countries, regions, and languages in which the app is available or localized. & ASO\_10 \\
LLM\_12 & Regular Updates and Maintenance & Frequency and quality of updates, including bug fixes, improvements, and new features. & ASO\_15 \\
LLM\_13 & Security and Privacy & Protection of user data and communications, including encryption, privacy policies, and compliance. & -- \\
LLM\_14 & User Engagement & How actively and frequently users interact with the app (e.g., session frequency, retention). & -- \\
LLM\_15 & User Interface and Experience & Ease of use, navigation, and overall user experience, including interface design and intuitiveness. & -- \\
LLM\_16 & User Ratings & Average rating from app store reviews, reflecting user satisfaction and perceived quality. & ASO\_04 \\
\bottomrule
\end{tabularx}
\end{table}

Concerning cross-configuration convergence (Step \step{6}), 10 of the 16 criteria are reported in all four configurations, covering the conceptual core of the taxonomy (e.g., \textit{Cost and Pricing}, \textit{Security and Privacy}, \textit{User Ratings}). The remaining six emerge from a subset of configurations: for instance, \textit{Accessibility Features} is contributed by the open-source cohort, whereas \textit{Feature Breadth and Richness} and \textit{Cross-Platform Synchronization} surface only in the proprietary cohort on the large-scale dataset. Per-criterion presence and frequency data across configurations are available in the replication package.

With respect to the information access ablation, enabling web search for the proprietary cohort on the small-scale dataset increased the volume of raw criteria (5,065 vs.\ 4,651 items from the same 640 queries) but did not expand the taxonomy: every criterion consolidated from the web-augmented configuration maps to an existing unified concept, with no web-exclusive criteria. 
The three unified criteria absent from the web-augmented results (\textit{Performance and Stability}, \textit{Accessibility Features}, \textit{Cross-Platform Synchronization}) originate from other knowledge-only configurations rather than from web retrieval. 
Full ablation data are reported in the replication package.

\subsection{Findings}
\label{sec:rq1-discussion}

Based on previous results, below we identify the main findings (F) and takeaways resulted from RQ\textsubscript{1}.

\begin{tcolorbox}[insightboxrq1]
\textbf{F1. LLMs report an extremely broad set of heterogeneous user-dependent and developer-dependent ranking criteria.} After filtering and consolidation, we distilled 16 LLM-based ranking criteria widely used across LLMs, model cohorts, and datasets.
\end{tcolorbox}

This synthesis introduces a significant generalization of the reasoning outcomes generated by LLMs, and hence cannot be interpreted as an exhaustive elicitation of the entire spectrum of possible ranking criteria. Nevertheless, it provides a simplified yet comprehensive perspective on the ranking criteria commonly adopted by multiple LLMs across heterogeneous app domains. 
Our replication package enables its reuse and extension to other software recommendation contexts, enhancing the transparency and explainability of LLM-based recommender systems (see Section~\ref{sec:method}, UC3).

\begin{tcolorbox}[insightboxrq1]
\textbf{F2. Limited alignment with ASO metrics, with substantial novelty beyond metadata-driven signals.} Only 6 of the 16 LLM-based ranking criteria align with traditional ASO metrics; the remaining 10 capture technical and experiential dimensions that ASO sources omit.
\end{tcolorbox}

The ASO taxonomy reflects a strong focus on developer-provided metadata (e.g., \textit{Name}, \textit{Price}, \textit{Content Rating}), whereas the 10 non-overlapping LLM criteria stress software quality and user experience dimensions such as \textit{User Interface and Experience} (LLM\_15) and \textit{Security and Privacy} (LLM\_13), functional capabilities such as \textit{Feature Breadth and Richness} (LLM\_06) and \textit{Integration with Other Services} (LLM\_07), and audience dynamics such as \textit{User Engagement} (LLM\_14). Conversely, notable ASO signals rooted in app metadata (e.g., app description, media content, developer identity) are rarely surfaced by LLMs as primary criteria. This complementary coverage could inform more nuanced decision-making for users (UC1) and strategic positioning for developers (UC2). 
Note that our ASO taxonomy is grounded in peer-reviewed literature and therefore excludes proprietary, non-public, or personalization-dependent industry signals (e.g., CTR, install velocity, or position bias), which are outside the scope of this study.

\begin{tcolorbox}[insightboxrq1]
\textbf{F3. Reliance on subjective and ambiguous indicators.} LLM-based criteria include subjective, user-centric indicators or ambiguously defined aspects.
\end{tcolorbox}

Criteria such as \textit{Security and Privacy} (LLM\_13) or \textit{User Interface and Experience} (LLM\_15) inherently contain subjective interpretations. Despite potential ambiguity, these criteria align with LLMs' strengths in natural language understanding and qualitative reasoning, suggesting that LLM-based recommenders might be particularly suitable for handling subjective evaluations and context-dependent user preferences.
A user (UC1) can express personal requirements and the LLM dynamically adapts recommendations, whereas ASO-based approaches rely on predefined criteria.

\begin{tcolorbox}[insightboxrq1]
\textbf{F4. Web search does not materially modify the elicited ranking criteria.} Enabling web search increases the lexical variability of reported criteria but yields no new ranking concepts, supporting its responsible exclusion from subsequent experiments.
\end{tcolorbox}

The information access ablation shows that all criteria consolidated under the web-augmented setting map to concepts already present in the knowledge-only taxonomy, with no web-exclusive criteria. This suggests that the reported ranking rationale is already driven by parametric knowledge. 
Given the substantial time, energy, and economic cost of web-augmented querying at scale, this result justifies conducting the remaining experiments in the knowledge-only setting (see Section~\ref{sec:information-access}).

\section{Recommendation Consistency (RQ\textsubscript{2})}
\label{sec:consistency}

\subsection{Design}

The consistency analysis of app recommendations is structured into internal consistency (Step~\step{7}), external consistency (Step~\step{8}), and an information access ablation (Step~\step{9}). This analysis is conducted on the large-scale feature dataset (see Section~\ref{sec:sample-study}), reusing the recommendations collected in RQ\textsubscript{1} (i.e., 10 runs for each feature and LLM, in the knowledge-only setting, across both model cohorts). For internal consistency, we focus on the degree of variability across multiple runs of the same LLM; for external consistency, we focus on the degree of variability between the 10 LLMs explored within this study.


\begin{itemize}

\item \textbf{Step \step{7} - Internal consistency analysis}. This analysis aims to assess how stable LLM recommendations are when prompted repeatedly under the same conditions. To this end, we analysed the consistency of app rankings within each individual LLM across the 10 runs for the same feature request, using \textit{Rank-Biased Overlap (RBO)}~\citep{Webber2010}. 
RBO measures the similarity between ranked lists with a controllable top-weighting bias ($p$), making it well-suited for recommendation scenarios where higher-ranked items dominate user attention. RBO ranges from 0 (completely disjoint top-ranked recommendations) to 1 (identical top-ranked recommendations), with intermediate values reflecting partial overlap weighted toward the top of the list.
RBO has been explicitly proposed to address the limitations of traditional rank correlation measures for indefinite, top-weighted ranked lists, and is widely adopted in information retrieval and recommender system evaluation. 
We set the top-weighting parameter to $p=0.9$, as a sensitivity analysis for $p \in \{0.8, 0.9, 1.0\}$ showed negligible variation in relative consistency scores. 
For each model--feature combination, we compared all pairwise combinations of runs and computed average RBO scores over the top-$k$ recommendations, with $k=20$ consistently with the recommendation depth used throughout our study (see Section~\ref{sec:elicitation}).

\item \textbf{Step \step{8} - External consistency analysis}. Complementarily, this analysis aims to assess to what extent different LLMs converge into similar ranking lists when given the same feature request. To this end, we compared app recommendations between all possible pairs of models 
for identical feature prompts, using the same metric and configuration ($k=20$) as in Step~\step{7}, computing average RBO scores across runs. 
This allows us to examine agreement patterns within and between cohorts (i.e., proprietary--proprietary, open--open, and proprietary--open pairs), assessing whether model openness relates to convergence in recommendation behaviour.

\item \textbf{Step \step{9} - Information access ablation}. We replicate the internal consistency analysis for the proprietary cohort on the small-scale dataset under both information access settings, comparing RBO scores with and without web search. 
This ablation assesses whether web search materially affects the stability of app recommendations, complementing the criteria-level ablation of RQ\textsubscript{1} and jointly supporting the knowledge-only design of RQ\textsubscript{3}.

\end{itemize}

\subsection{Results}

\begin{figure}[pos=htb]
    \centering
    \includegraphics[width=\columnwidth]{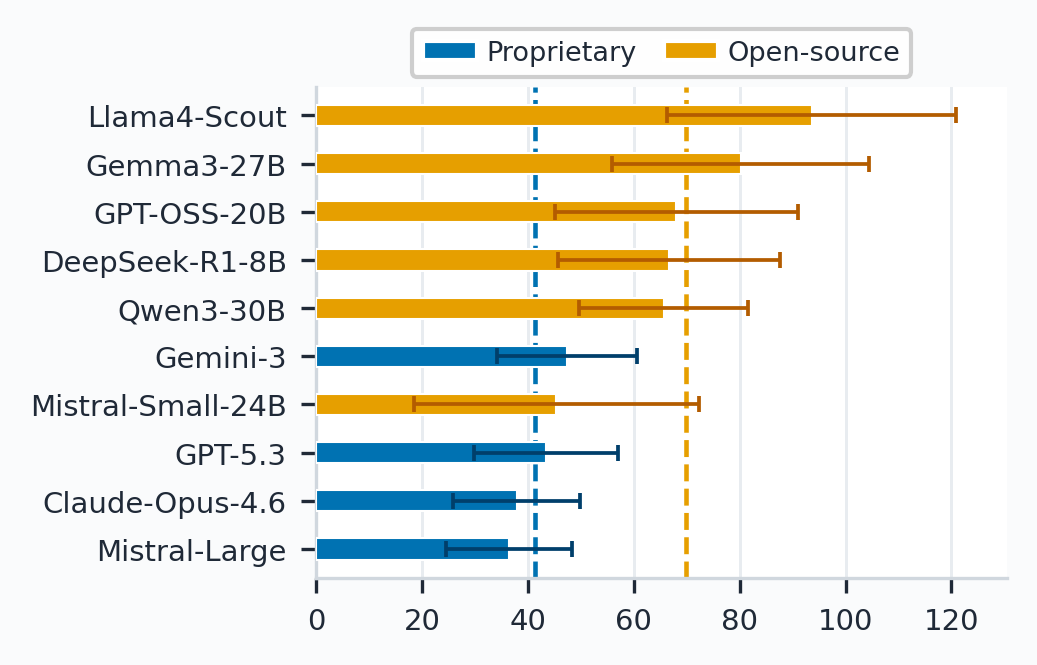}
    \caption{Average number of distinct apps recommended per model across 10 runs per feature (large-scale dataset, $k=20$).}
    \label{fig:distinct-apps}
\end{figure}

We begin with a high-level view of output variability. Figure~\ref{fig:distinct-apps} shows the average number of distinct apps recommended by each LLM across the 10 runs per feature (with $k=20$, a fully stable model would produce 20 distinct apps; a totally unstable one 200). Proprietary models produce markedly more concentrated outputs --- \texttt{mistral-large-2512} (36.4) and \texttt{claude-opus-4-6} (37.9) recommend the fewest distinct apps --- whereas open-source models exhibit substantially higher dispersion, peaking with \texttt{llama4:scout} (93.6, i.e., nearly five distinct apps per available ranking position). This dispersion gap anticipates the cohort-level differences observed in the consistency analysis.

\begin{figure}[pos=htb]
    \centering
    \includegraphics[width=\columnwidth]{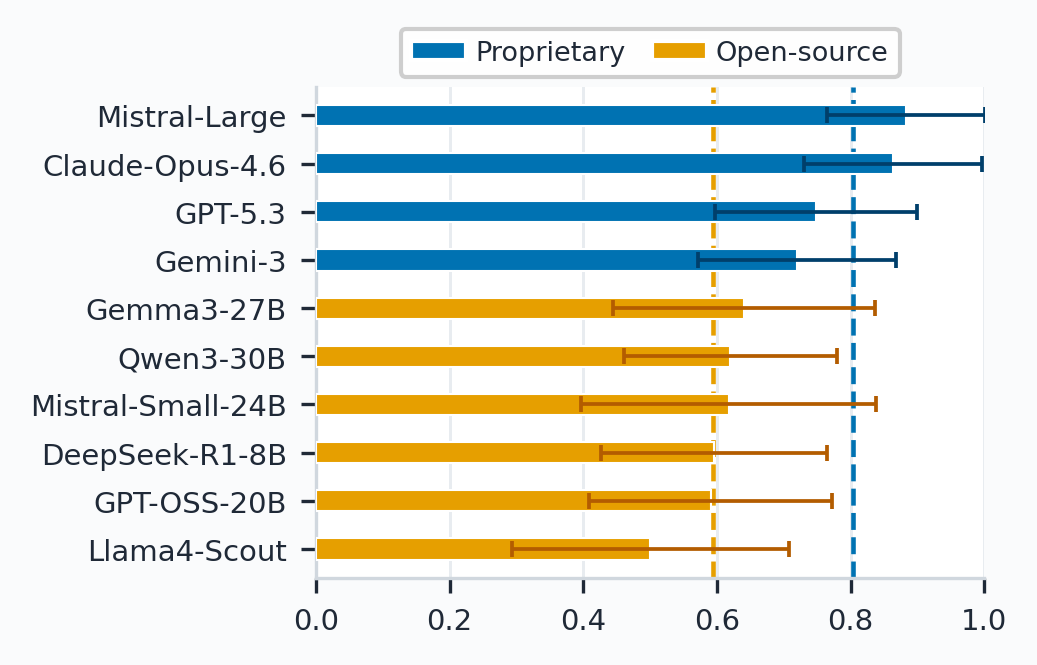}
    \caption{Internal consistency (mean RBO, $k=20$, $p=0.9$) per model, grouped by cohort.}
    \label{fig:internal-consistency}
\end{figure}

Concerning internal consistency (Step~\step{7}), Figure~\ref{fig:internal-consistency} reports the average RBO per model across all features and run pairs. The proprietary cohort is systematically more stable than the open-source cohort (mean RBO $0.804$ vs.\ $0.594$ respectively, $0.678$ combined). At the model level, \texttt{mistral-large-2512} ($0.883$) and \texttt{claude-opus-4-6} ($0.864$) lead by a clear margin, followed by \texttt{gpt-5.3-chat} ($0.748$) and \texttt{gemini-3-flash} ($0.719$); all open-source models score below the least consistent proprietary model, ranging from \texttt{gemma3:27b} ($0.640$) down to \texttt{llama4:scout} ($0.500$). 
Internal consistency is also strongly feature-dependent.
As an example, features related to common categories such as music and messaging yield highly stable recommendations (e.g., \textit{Stream music} = $0.905$, \textit{Create playlists} = $0.884$), whereas less generic features such as customization and finance exhibit low stability across all models (e.g., \textit{Customize themes} = $0.388$, \textit{Change fonts} = $0.298$). Even the most stable models are affected: \texttt{claude-opus-4-6} drops to $0.326$ for \textit{Change fonts}, and \texttt{gpt-5.3-chat} to $0.282$ for \textit{Financial calculators}. 
This suggests that stability is jointly determined by the model and by the maturity or fragmentation of the app domain being queried, with niche utility domains lacking a consolidated set of dominant apps.

\begin{figure}[pos=htb]
    \centering
    \includegraphics[width=\columnwidth]{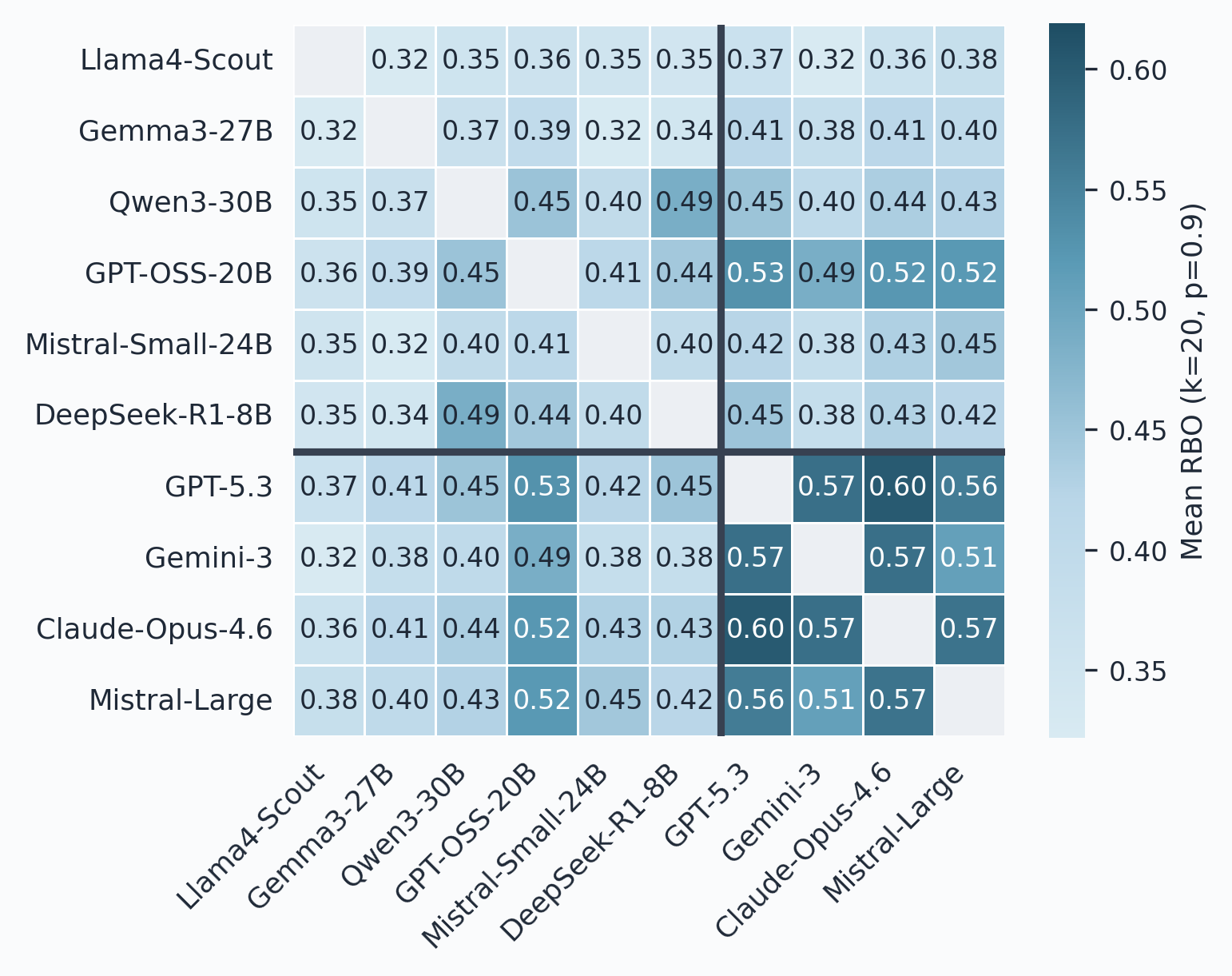}
    \caption{External consistency of app recommendations: pairwise mean RBO ($k=20$, $p=0.9$) between all model pairs, grouped by cohort.}
    \label{fig:external-consistency}
\end{figure}

Concerning external consistency (Step~\step{8}), Figure~\ref{fig:external-consistency} presents the pairwise RBO matrix between all 10 models, aggregated across features and runs. Cross-model agreement (overall mean $0.430$) is substantially lower than within-model stability ($0.678$), confirming that different LLMs converge only partially on the same ranked recommendations even under identical prompting conditions. The cohort structure is clearly visible: proprietary--proprietary pairs agree the most (mean $0.565$), followed by cross-cohort pairs ($0.424$) and open--open pairs ($0.384$). The highest agreement is observed for \texttt{gpt-5.3-chat}--\texttt{claude-opus-4-6} ($0.602$), and the lowest for \texttt{llama4:scout}--\texttt{gemma3:27b} ($0.322$).
Notably, \texttt{llama4:scout} (the model with the lowest internal consistency) is involved in the weakest cross-model agreements, indicating that instability propagates to divergence. Feature-level effects mirror the internal analysis but with amplified extremes: agreement is high for consolidated domains (\textit{Stream music} = $0.789$, \textit{Create playlists} = $0.769$) and near-zero for fragmented ones (\textit{Financial calculators} = $0.043$, \textit{Change fonts} = $0.045$), where each model effectively surfaces a different app ecosystem.

\begin{figure}[pos=htb]
  \centering
  \includegraphics[width=\columnwidth]{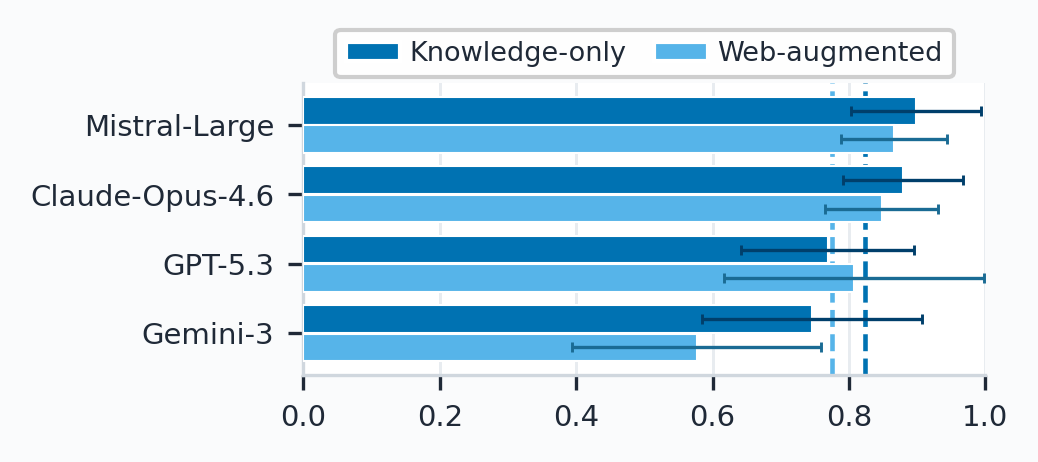}
  \caption{Information access ablation: internal consistency (mean RBO, $k=20$, $p=0.9$) of the proprietary cohort on the small-scale dataset, with and without web search.}
\label{fig:ablation-internal-rbo}
\end{figure}

Concerning the information access ablation (Step~\step{9}), Figure~\ref{fig:ablation-internal-rbo} compares the internal consistency of the proprietary cohort with and without web search on the small-scale dataset. Enabling web search yields a small reduction in internal consistency (mean RBO $0.775$ vs.\ $0.823$; paired $t$-test over model--feature cells: $t=2.38$, $p=0.021$, Cohen's $d=0.30$). The effect is not uniform across models: \texttt{gemini-3-flash} accounts for most of the drop ($0.577$ vs.\ $0.746$), whereas the remaining models shift marginally in either direction. Overall, web search introduces moderate additional variability without altering the relative stability ranking of the models, complementing the criteria-level ablation of RQ\textsubscript{1} and jointly supporting the knowledge-only design of the remaining experiments (see Section~\ref{sec:information-access}).

\subsection{Findings}
\label{sec:rq2-discussion}

\begin{tcolorbox}[insightboxrq2]
\textbf{F5. Proprietary models generate substantially more stable recommendations than open-source models.} Internal consistency shows a marked cohort gap, with every open-source model scoring below the least consistent proprietary model.
\end{tcolorbox}

Internal consistency analyses reveal a systematic stability gap between cohorts, 
consistent with the higher output dispersion of open-source models (Figure~\ref{fig:distinct-apps}). Several factors may contribute to this gap, including differences in model scale imposed by local deployment constraints, training and alignment procedures, and default decoding configurations. From a practical standpoint, this implies that recommendation stability is currently a differentiating property of proprietary systems, and that reproducibility-sensitive applications relying on locally deployed models should account for higher output variability, e.g., through aggregation across multiple runs. 
At the same time, stability might not be inherently desirable: more variable models may surface a broader candidate space, a property we revisit in RQ\textsubscript{3}.

\begin{tcolorbox}[insightboxrq2]
\textbf{F6. External consistency indicates substantial divergence between models, revealing inherent differences in ranking logic and prioritization across LLMs.} Cross-model agreement follows a clear cohort structure, underscoring the practical importance of model selection according to the intended recommendation scenario.
\end{tcolorbox}

External consistency analyses emphasize significant differences among LLMs regarding their recommendation strategies and internal prioritization of app qualities, with cross-model agreement (mean RBO $0.430$) substantially below within-model stability ($0.678$). Pairwise comparisons reveal that proprietary models converge most among themselves, open-source pairs diverge the most, and cross-cohort agreement lies in between. 
This suggests partially shared ranking logic among leading proprietary systems 
contrasted with more idiosyncratic behaviour among open-source models. 
This reinforces the importance of model selection tailored to specific stakeholder priorities, whether consistency, interpretability, or diversity. 
From an evaluation perspective, the methodology introduced in this study provides a reusable mechanism for generating consistency-based baselines, enabling future comparisons across evolving LLMs, app stores, classical recommenders, or hybrid approaches.

\begin{tcolorbox}[insightboxrq2]
\textbf{F7. Consistency is strongly domain-dependent, both within and across models.} Features targeting consolidated app domains yield stable, convergent recommendations, whereas niche or fragmented domains produce highly divergent outputs across models.
\end{tcolorbox}

Features associated with mature app ecosystems dominated by a few well-established apps (e.g., music streaming, messaging) exhibit high internal and external consistency, while features in fragmented domains (e.g., financial utilities, interface customization) drop to near-zero cross-model agreement, with each model effectively surfacing a different set of candidate apps. This pattern affects even the most stable models, indicating that domain structure 
bounds achievable consistency. 
For users (UC1), this implies that LLM recommendations are most trustworthy in consolidated domains. 
For system designers (UC3), fragmented domains are precisely where consistency-enhancing mechanisms, such as explicit ranking criteria, may add the most value. 

\begin{tcolorbox}[insightboxrq2]
\textbf{F8. Web search does not materially alter recommendation consistency.} Enabling web search introduces a small, model-dependent increase in variability without changing the relative stability of models.
\end{tcolorbox}

The information access ablation shows a small overall reduction in internal consistency under the web-augmented setting (Cohen's $d=0.30$), concentrated almost entirely in a single model, while the stability ranking of the proprietary cohort remains unchanged. Together with the criteria-level ablation (F4), this indicates that our two objects of analysis (i.e., ranking criteria and  consistency) are not materially affected by web retrieval, supporting the knowledge-only design of the remaining experiments as a responsible use of computational resources. 
Nevertheless, web search may well affect which specific apps are recommended and their factual currency, which our study deliberately does not assess.
\section{Ranking Criteria Impact (RQ\textsubscript{3})}
\label{sec:alignment}

\subsection{Design}

The ranking criteria convergence analysis is structured into two main stages. First, we conduct a systematic prompt-based experiment (similar to RQ\textsubscript{1}) to retrieve mobile app recommendations conditioned on specific ranking criteria (Step~\step{10}). We refer to these as \textit{guided} app recommendations, in contrast with the \textit{blind} recommendations collected in RQ\textsubscript{1}, which were generated without conditioning on any ranking criterion. Second, we analyse the effect of this conditioning on recommendation consistency (Step~\step{11}), using the consistency of blind recommendations on the same dataset (see Section~\ref{sec:consistency}) as a baseline. Our working hypothesis is that an explicit ranking criterion acts as a shared reference that reduces model-specific interpretation, increasing both the determinism of individual models and the convergence between models.



We used the same system prompt as in RQ\textsubscript{1} to ensure consistency across experiments and minimize variability. For the user prompt\footnote{The example in grey text is not included in the actual prompt; it is shown in the manuscript for explanatory purposes.}, we made a slight adaptation of the one used in RQ\textsubscript{1} by appending the \texttt{ranking\_criteria} as an extension to the feature-based query. Here, \texttt{ranking\_criteria} refers to the JSON representation of a given ranking criterion from the taxonomy elicited in RQ\textsubscript{1} (see Table~\ref{tab:llm-aso-mapping}).

\begin{tcolorbox}[
colback=gray!5!white,
colbacktitle=purple!50!black,
coltitle=white,
title=\textbf{User prompt (RQ\textsubscript{3})},
boxrule=0pt,
left=0.5mm, right=0.5mm, top=0.5mm, bottom=0.5mm,
sharp corners,
enhanced
]
Recommend \texttt{\{k\}} apps to \texttt{\{feature\}} based on the following ranking criteria: \texttt{\{ranking\_criteria\}}. \\
\textcolor{gray}{\# E.g., Recommend 20 apps to build photo collages based on the following ranking criteria: \{"n": "Feature Breadth and Richness", "d": "The range, depth, and quality of features offered by the app, including core functionality and optional capabilities."\}.}
\end{tcolorbox}

Below, we summarize the details of the experiment and the corresponding analysis.

\begin{itemize}
    \item \textbf{Step \step{10} - Guided LLM app recommendations.} Using the same configuration and service as in Step \step{3}, we set \texttt{k}=20 and prompted 5 independent runs for each \texttt{feature}-\texttt{ranking\_criteria} combination. Hence, each query instructs the LLM to use the given ranking criterion to select and rank the set of mobile apps for the given \texttt{feature}. Using the small dataset, this led to 16 features $\times$ 16 ranking criteria $\times$ 5 runs $\times$ 10 LLMs $=$ 12,800 queries. App recommendations were consolidated using the same structure as in RQ\textsubscript{1}.
    \item \textbf{Step \step{11} - Convergence analysis.} We assess the effect of criteria conditioning from two complementary perspectives, reusing the consistency metric and configuration of RQ\textsubscript{2} (RBO, $k=20$, $p=0.9$). First, \textit{internal convergence}: for each model and ranking criterion, we compute the internal consistency of guided recommendations (pairwise RBO across the 5 runs, averaged over features) and compare it against the internal consistency of blind recommendations for the same model on the same dataset. 
    Second, \textit{external convergence}: for each ranking criterion, we compute the external consistency of guided recommendations (pairwise RBO across all model pairs) and compare it against the external consistency baseline of blind recommendations. 
    Additionally, we compute the RBO between guided and blind recommendations per criterion and model, characterizing how far conditioning displaces recommendations from their blind baseline.
\end{itemize}

\subsection{Results}

\begin{figure*}[pos=htb]
    \centering
    \includegraphics[width=\textwidth]{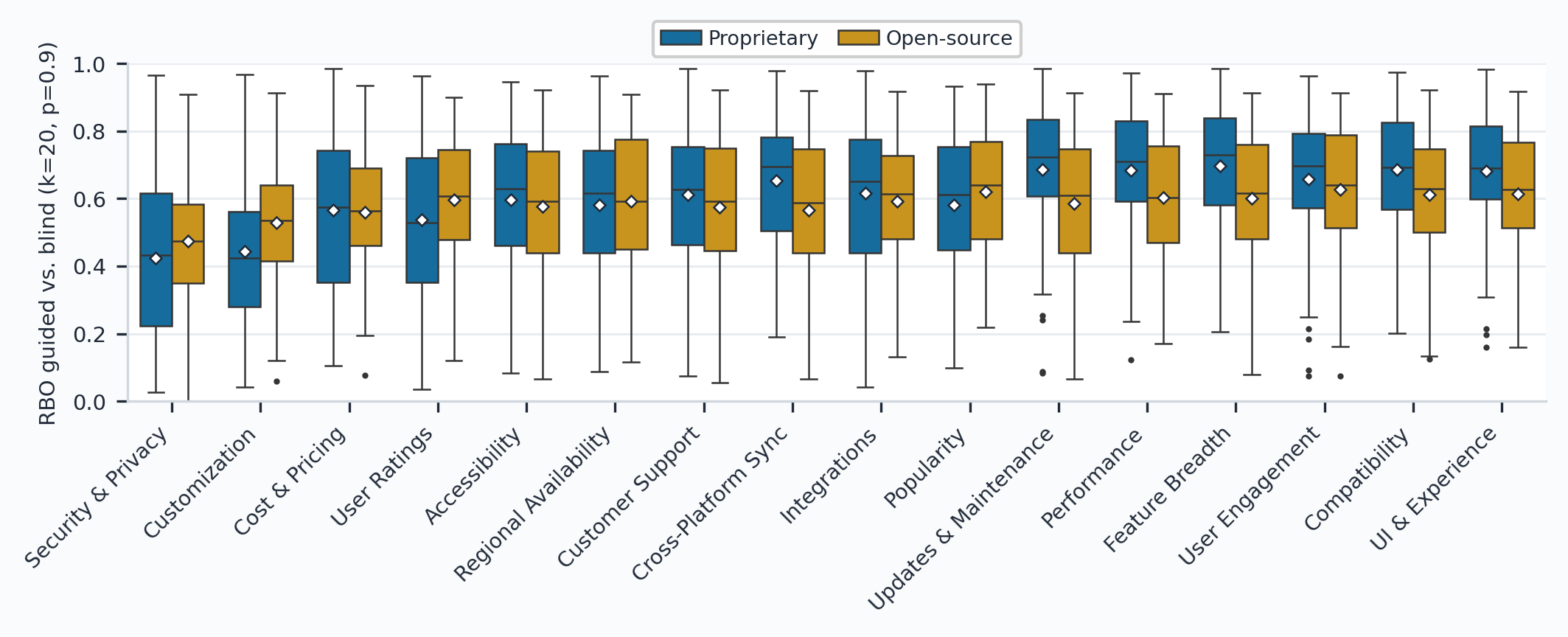}
    \caption{Sensitivity to ranking criteria (RQ\textsubscript{3}): distribution of RBO ($k=20$, $p=0.9$) between guided and blind recommendations per ranking criterion. Criteria are ordered from most (left) to least (right) displacing; white diamonds mark distribution means.}
    \label{fig:rq3-displacement}
\end{figure*}

Figure~\ref{fig:rq3-displacement} displays the distribution of RBO values between guided and blind recommendations per ranking criterion and model cohort, across all model~$\times$~feature cells. 
With an overall average RBO of $0.592$ across all models and criteria, the results indicate moderate overlap between guided and blind recommendations. 
This value is substantially lower than the average internal consistency of blind recommendations ($0.678$; see Section~\ref{sec:consistency}), confirming that the inclusion of ranking criteria alters model recommendations well beyond run-to-run variability.

Sensitivity is clearly stratified by model: \texttt{mistral-large-2512} exhibits the highest overlap between guided and blind recommendations ($0.711$), followed by \texttt{mistral-small3.1:24b} ($0.650$) and \texttt{claude-opus-4-6} ($0.632$), suggesting that these models either inherently consider multiple ranking factors without explicit guidance or are less responsive to criteria-conditioned prompts. At the other end, \texttt{gemini-3-flash} ($0.475$), \texttt{llama4:scout} ($0.498$) and \texttt{gpt-oss:20b} ($0.533$) restructure their rankings most aggressively. On aggregate, open-source models are slightly more sensitive than proprietary ones ($0.583$ vs.\ $0.607$).

The degree of displacement is also strongly criterion-dependent (Table~\ref{tab:rq3-per-criterion}). \textit{Security and Privacy} triggers the largest changes ($0.454$), followed by \textit{Customization Options} ($0.494$), \textit{Cost and Pricing} ($0.562$) and \textit{User Ratings} ($0.573$), whereas \textit{User Interface and Experience} and \textit{Platform and Device Compatibility} (both $0.642$), \textit{User Engagement} and \textit{Feature Breadth and Richness} (both $0.640$) remain closest to the blind baseline. The most displacing criteria also exhibit the widest spread across models (std $\geq 0.094$ for \textit{Security and Privacy}, \textit{Customization Options}, \textit{User Ratings}, \textit{Customer Support} and \textit{Cost and Pricing}), indicating uneven responsiveness across LLMs, while \textit{Regional and Geographical Availability} shows the most homogeneous behaviour (std $= 0.045$).

Conditioning systematically surfaces criterion-relevant apps that are absent from blind recommendations. 
For instance, under \textit{Security and Privacy}, Threema\footnote{\url{https://threema.com/}} enters the top-5 (median rank 2) for the feature \textit{keeping up with friends} for five different models (\texttt{claude-opus-4-6}, \texttt{gemini-3-flash}, \texttt{gpt-oss:20b}, \texttt{mistral-small3.1:24b}, \texttt{qwen3:30b-a3b}) despite being absent from all their blind top-20 lists, and Wire\footnote{\url{https://wire.com/}} and Proton Mail\footnote{\url{https://proton.me/mail}} exhibit similar behaviour for messaging and file-sharing features. 
Under \textit{Popularity and Market Reach}, three of the four proprietary models rank Instagram among the top-5 for \textit{build photo collage}, an app none of them recommends for that feature in the blind setting.
This pattern is repeated across multiple features, apps, ranking criteria, and LLMs\footnote{We limit the scope of analysis included in this paper to overall and aggregated results. Detailed data per feature, model, and ranking criterion are available in the replication package.}.

\begin{table*}[pos=htb]
\centering
\caption{Sensitivity and convergence of guided recommendations per ranking criterion (RQ\textsubscript{3}). G--B: mean RBO between guided and blind recommendations; $\Delta$Int.: mean change in internal RBO (guided $-$ blind) across models; Ext.: mean external RBO of guided recommendations across all 45 model pairs; $\Delta$Ext.: difference against the blind external baseline; $p_{\text{Holm}}$ and $d_z$: Holm-corrected Wilcoxon signed-rank $p$-value and Cohen's effect size, paired per feature. All RBO values with $k=20$, $p=0.9$.}
\label{tab:rq3-per-criterion}
\begin{tabular*}{\textwidth}{@{\extracolsep{\fill}}lrrrrrr@{}}
\toprule
\textbf{Ranking criterion} & \textbf{G--B} & \textbf{$\Delta$Int.} & \textbf{Ext.} & \textbf{$\Delta$Ext.} & \textbf{$p_{\text{Holm}}$} & \textbf{$d_z$} \\
\midrule
Popularity and Market Reach            & 0.605 & $+0.020$ & 0.477 & $+0.004$ & 1.000 & $+0.05$ \\
User Engagement                        & 0.640 & $+0.010$ & 0.474 & $+0.001$ & 1.000 & $+0.02$ \\
Platform and Device Compatibility      & 0.642 & $-0.007$ & 0.455 & $-0.017$ & 0.523 & $-0.33$ \\
Cross-Platform Synchronization         & 0.601 & $-0.005$ & 0.451 & $-0.021$ & 0.523 & $-0.30$ \\
Regional and Geographical Availability & 0.588 & $-0.024$ & 0.443 & $-0.029$ & 0.500 & $-0.56$ \\
Integration with Other Services        & 0.603 & $-0.019$ & 0.439 & $-0.034$ & 0.523 & $-0.40$ \\
Accessibility Features                 & 0.584 & $-0.046$ & 0.437 & $-0.036$ & 0.235 & $-0.61$ \\
Performance and Stability              & 0.635 & $-0.029$ & 0.436 & $-0.036$ & 0.046 & $-0.82$ \\
Feature Breadth and Richness           & 0.640 & $-0.028$ & 0.434 & $-0.039$ & 0.092 & $-0.78$ \\
User Interface and Experience          & 0.642 & $-0.028$ & 0.428 & $-0.045$ & 0.002 & $-1.14$ \\
Regular Updates and Maintenance        & 0.627 & $-0.041$ & 0.428 & $-0.045$ & 0.099 & $-0.75$ \\
Customer Support                       & 0.590 & $-0.051$ & 0.423 & $-0.049$ & 0.146 & $-0.64$ \\
Cost and Pricing                       & 0.562 & $-0.062$ & 0.364 & $-0.109$ & $<$0.001 & $-1.31$ \\
User Ratings                           & 0.573 & $-0.087$ & 0.353 & $-0.119$ & $<$0.001 & $-1.69$ \\
Security and Privacy                   & 0.454 & $-0.112$ & 0.340 & $-0.133$ & $<$0.001 & $-1.44$ \\
Customization Options                  & 0.494 & $-0.095$ & 0.332 & $-0.141$ & 0.007 & $-1.10$ \\
\bottomrule
\end{tabular*}
\end{table*}

\begin{figure*}[pos=htb]
    \centering
    \includegraphics[width=\textwidth]{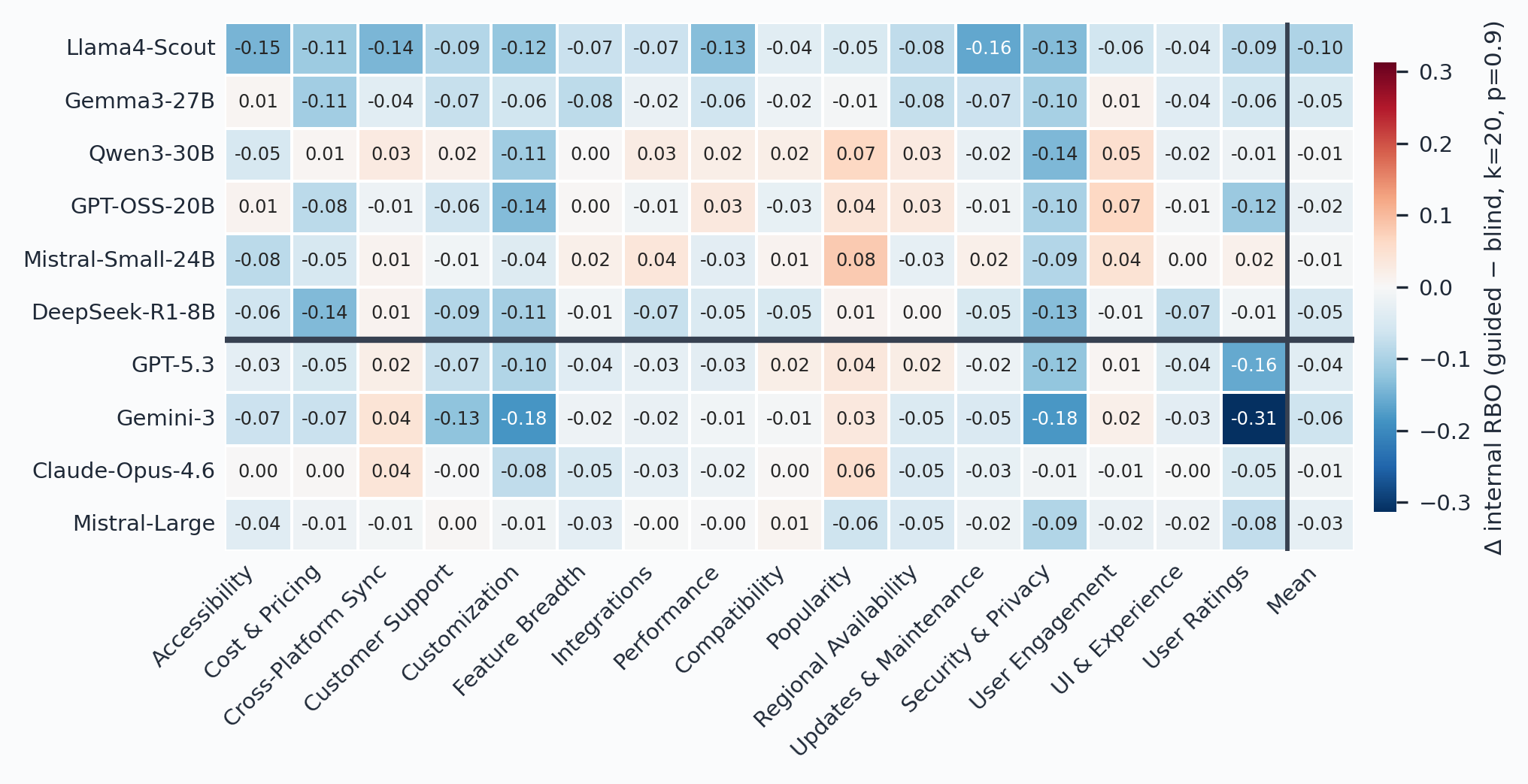}
    \caption{Internal convergence of guided recommendations (RQ\textsubscript{3}): difference in internal RBO ($k=20$, $p=0.9$) between guided and blind recommendations, per model and ranking criterion. Blue cells indicate reduced determinism under criteria conditioning; red cells indicate increased determinism. The horizontal line separates open-source (top) from proprietary (bottom) models; the rightmost column reports the per-model average across criteria.}
    \label{fig:rq3-internal-delta}
\end{figure*}

Having established that criteria conditioning effectively steers recommendations, we examine whether it also stabilizes them. Figure~\ref{fig:rq3-internal-delta} reports, for each model and ranking criterion, the difference in internal consistency between guided and blind recommendations ($\Delta$ internal RBO), computed per feature and averaged over the 16 features. Contrary to our working hypothesis, conditioning does not increase model determinism: the average delta across all model~$\times$~criterion cells is $-0.038$, 
and no model becomes more deterministic on average. 
Degradation is mildest for \texttt{qwen3:30b-a3b} and \texttt{mistral-small3.1:24b} (mean $\Delta = -0.006$) and \texttt{claude-opus-4-6} ($-0.014$), and most pronounced for \texttt{llama4:scout} ($-0.096$, negative under all 16 criteria), \texttt{gemini-3-flash} ($-0.064$) and \texttt{deepseek-r1:8b} ($-0.052$).

At the criterion level, only \textit{Popularity and Market Reach} ($+0.020$) and \textit{User Engagement} ($+0.010$) yield a positive average delta across models (Table~\ref{tab:rq3-per-criterion}). All remaining criteria reduce run-to-run consistency, with the strongest degradation concentrated in \textit{Security and Privacy} ($-0.112$), \textit{Customization Options} ($-0.095$), \textit{User Ratings} ($-0.087$) and \textit{Cost and Pricing} ($-0.062$), which is the most displacing criteria identified above. The single largest drop is observed for \texttt{gemini-3-flash} under \textit{User Ratings} ($-0.313$), whose blind rankings for a given feature are nearly deterministic but whose guided runs disperse over a much broader set of candidate apps.

\begin{figure}[pos=htb]
    \centering
    \includegraphics[width=\columnwidth]{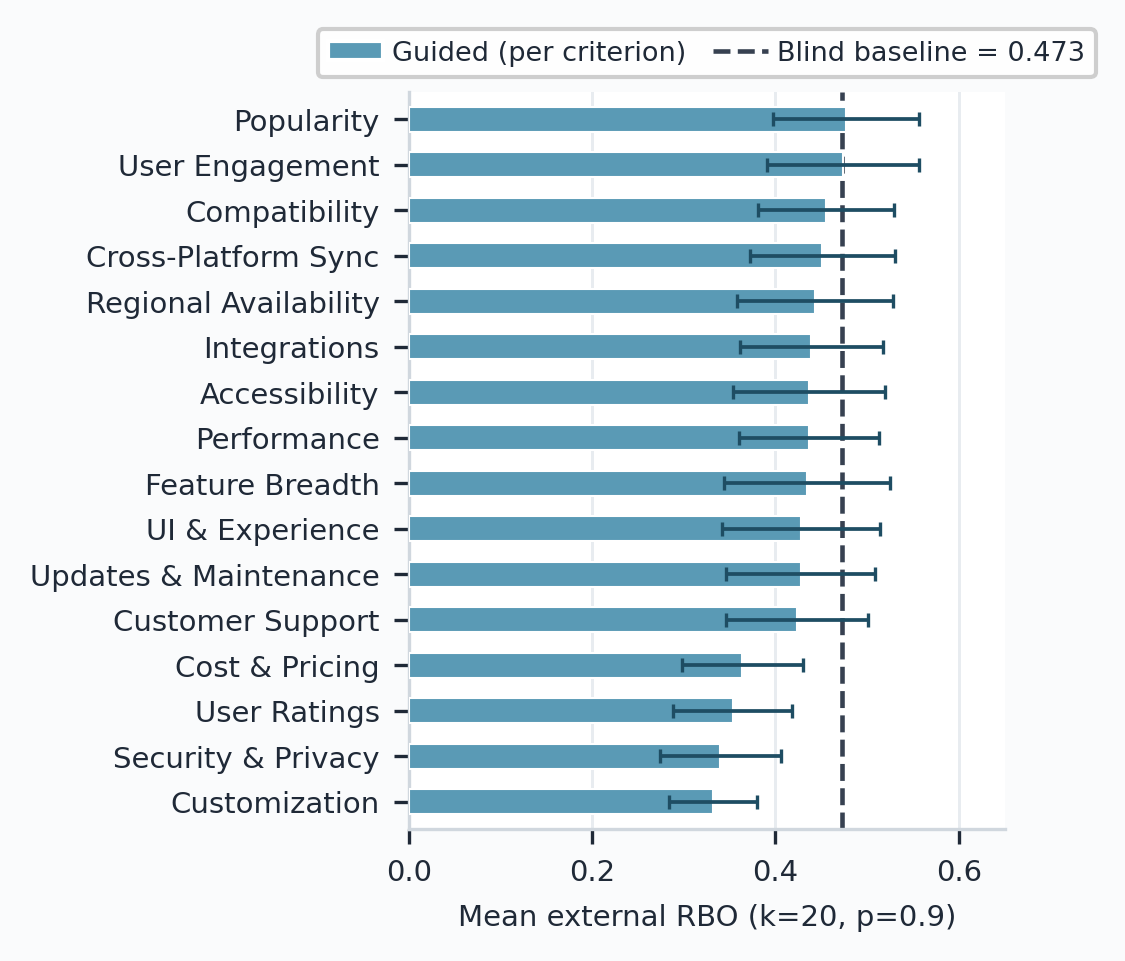}
    \caption{External convergence of guided recommendations per ranking criterion (RQ\textsubscript{3}): mean external RBO ($k=20$, $p=0.9$) across all 45 model pairs, with 95\% confidence intervals across features. The dashed line marks the external consistency baseline of blind recommendations.}
    \label{fig:rq3-external}
\end{figure}

Figure~\ref{fig:rq3-external} shows the external consistency of guided recommendations per ranking criterion (mean pairwise RBO across all 45 model pairs, averaged over features and runs), against the external consistency baseline of blind recommendations. 
The convergence hypothesis is again not supported: 14 of the 16 criteria fall below the blind baseline. Only \textit{Popularity and Market Reach} ($0.477$, $\Delta = +0.004$) and \textit{User Engagement} ($0.474$, $+0.001$) match the baseline (the same two criteria that preserve internal determinism) and neither difference is statistically significant. 
To assess significance, we paired guided and blind external consistency per feature (16 pairs per criterion) and applied a two-sided Wilcoxon signed-rank test with Holm correction across the 16 criteria, reporting Cohen's $d_z$ as effect size (Table~\ref{tab:rq3-per-criterion}).
Six criteria significantly reduce cross-model agreement ($p_{\text{Holm}} < 0.05$) with large effect sizes: \textit{Customization Options} ($0.332$, $\Delta = -0.141$, $d_z = -1.10$), \textit{Security and Privacy} ($0.340$, $-0.133$, $d_z = -1.44$), \textit{User Ratings} ($0.353$, $-0.119$, $d_z = -1.69$), \textit{Cost and Pricing} ($0.364$, $-0.109$, $d_z = -1.31$), \textit{User Interface and Experience} ($0.428$, $-0.045$, $d_z = -1.14$) and \textit{Performance and Stability} ($0.436$, $-0.036$, $d_z = -0.82$). The remaining criteria show moderate, non-significant reductions ($\Delta$ between $-0.017$ and $-0.049$).

\begin{figure*}[pos=htb]
    \centering
    \includegraphics[width=\textwidth]{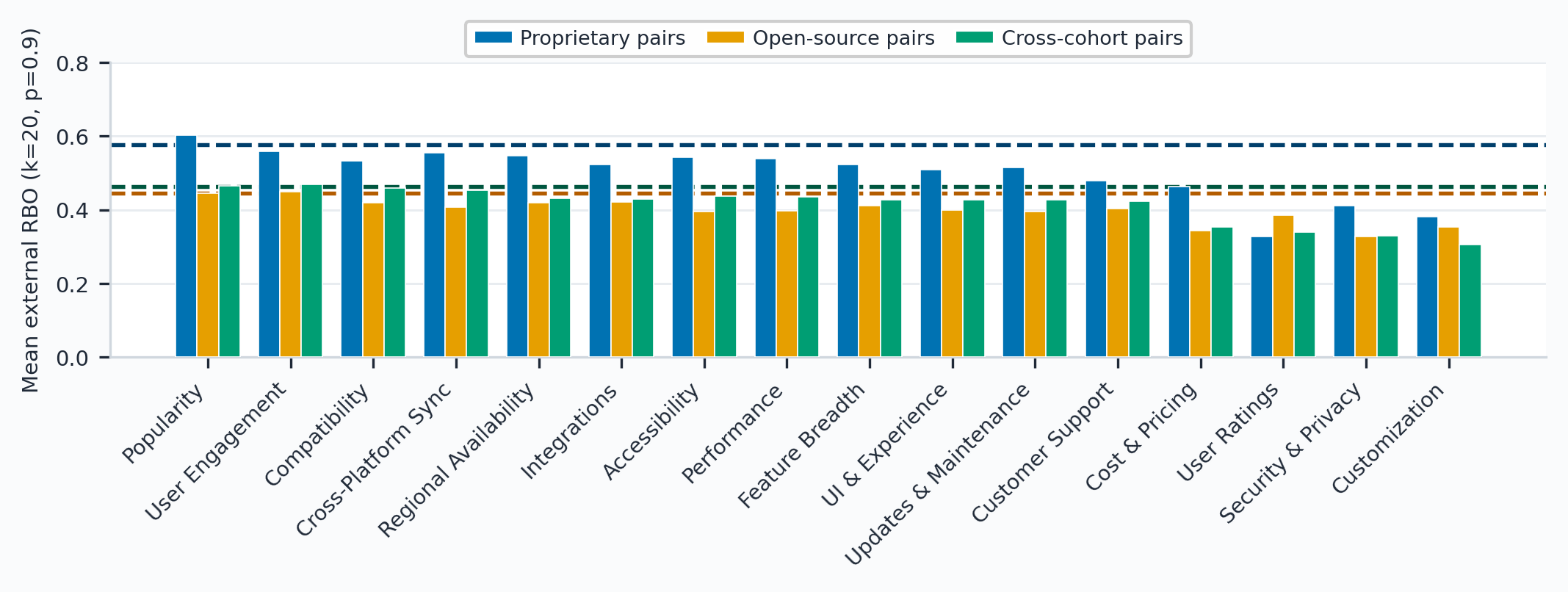}
    \caption{External convergence of guided recommendations per ranking criterion and model cohort (RQ\textsubscript{3}): mean external RBO ($k=20$, $p=0.9$) within proprietary pairs, within open-source pairs, and across cohorts. Dashed lines mark the corresponding blind baselines per group.}
    \label{fig:rq3-cohort}
\end{figure*}

Breaking down external convergence by cohort (Figure~\ref{fig:rq3-cohort}), the ordering observed for blind recommendations in RQ\textsubscript{2} (proprietary pairs $0.577$ $>$ cross-cohort pairs $0.463$ $>$ open-source pairs $0.444$) is preserved under conditioning for 15 of the 16 criteria: proprietary models remain the most mutually consistent cohort, although all three groups shift downwards roughly in parallel. The exception is \textit{User Ratings}, under which proprietary agreement collapses to $0.327$ (below both open-source ($0.385$) and cross-cohort ($0.340$) agreement) making it the only criterion for which proprietary models diverge the most.

Taken together, the three analyses reveal an inverse relation between sensitivity and convergence: the criteria that displace recommendations the most are also those that most degrade internal and external consistency, while the two criteria that preserve consistency (\textit{Popularity and Market Reach}, \textit{User Engagement}) are among those that displace recommendations the least. Explicit ranking criteria thus act as an effective steering mechanism, but not as a convergence mechanism.

\subsection{Findings}
\label{sec:rq3-discussion}

\begin{tcolorbox}[insightboxrq3]
\textbf{F9. Explicit ranking criteria effectively steer recommendations, substantially restructuring rankings and surfacing apps absent from unconditioned outputs.} Conditioning alters recommendations well beyond run-to-run variability, with subjective and user-centric criteria showing the strongest impacts.
\end{tcolorbox}

Guided recommendations retain only moderate overlap with their blind counterparts, 
confirming that conditioning triggers a genuine restructuring of rankings rather than noise-level reshuffling, and that part of the displaced positions are occupied by apps that never appear in the blind setting. Anecdotal inspection suggests that these newly surfaced apps are often plausible responses to the requested criterion, 
although our experimental design does not allow us to systematically verify such alignment. 
However, this steering capacity demonstrates the potential of generative recommenders to support diversity and novelty in app discovery, which is particularly relevant for users seeking personalized or niche solutions (UC1), and for developers aiming to enhance visibility through strategic alignment with emergent ranking dimensions (UC2). 
Impact is strongest for qualitative, user-centric criteria (\textit{Security and Privacy}, \textit{Customization Options}) and weakest for criteria adjacent to the models' default ranking behaviour (\textit{User Interface and Experience}, \textit{Platform and Device Compatibility}, \textit{User Engagement}). Sensitivity is also model-dependent: \texttt{gemini-3-flash} and \texttt{llama4:scout} restructure their rankings most aggressively, whereas \texttt{mistral-large-2512} and \texttt{claude-opus-4-6} remain closest to their unconditioned behaviour, suggesting they either inherently weigh multiple ranking factors or are less responsive to criteria-conditioned instructions.

\begin{tcolorbox}[insightboxrq3]
\textbf{F10. Contrary to our convergence hypothesis, criteria conditioning does not stabilize recommendations: it reduces both within-model determinism and cross-model agreement.} An explicit ranking criterion does not act as a shared reference that aligns models; it amplifies model-specific interpretation.
\end{tcolorbox}

If models possessed reliable, criterion-specific knowledge, conditioning should reduce the space of valid answers and make outputs more reproducible. We observe the opposite: internal consistency decreases for every model on average, 
and external consistency falls below the blind baseline for 14 of 16 criteria, significantly for six of them with large effect sizes. The only criteria that preserve consistency (i.e., \textit{Popularity and Market Reach} and \textit{User Engagement}) are precisely those that restate the popularity-oriented behaviour models exhibit by default. Part of the run-to-run variability may be amplified by default temperature settings, but temperature alone cannot explain the cross-model divergence, nor its systematic dependence on the criterion. For practitioners, this implies that criteria-conditioned prompting cannot be used as a mechanism to obtain more predictable or interoperable recommendations across LLM providers; achieving that would require more structured prompting frameworks or hybrid approaches that combine generative capabilities with explicit rule-based or retrieval-enhanced mechanisms grounding each criterion in verifiable data.

\begin{tcolorbox}[insightboxrq3]
\textbf{F11. Ranking criteria act as a steering mechanism, not a convergence mechanism: the further a criterion departs from the models' shared popularity prior, the more it destabilizes recommendations.} Cross-model agreement in blind settings reflects a shared popularity-driven default rather than shared criterion-specific knowledge.
\end{tcolorbox}

Across the three analyses, sensitivity and convergence are inversely related: the criteria that displace recommendations the most (\textit{Security and Privacy}, \textit{Customization Options}, \textit{Cost and Pricing}, \textit{User Ratings}) are also those that most degrade internal and external consistency, while the least displacing criteria preserve it. This pattern suggests that the blind ranking constitutes a stable attractor --- essentially the popularity prior shared across models, consistent with the market concentration observed in RQ\textsubscript{2} --- and that conditioning on criteria requiring knowledge models do not reliably possess (e.g., factual pricing, up-to-date ratings, security posture) pushes generation into sparser regions of their knowledge, where each run and each model improvises differently. The cohort analysis reinforces this reading: proprietary models remain the most mutually consistent group under nearly all criteria, yet their agreement collapses below that of open-source models under \textit{User Ratings}, the criterion most dependent on volatile, store-specific factual data. Models therefore respond strongly to the presence of a criterion (F9), but lack a shared, grounded basis for ranking by it (F10).

Notice that these experiments do not assess the correctness of app recommendations with respect to ranking criteria. Instead, our focus is on measuring the sensitivity and convergence of LLM outputs when exposed to different ranking prompts. Evaluating whether the recommended apps are objectively aligned with the intended criterion (e.g., whether Threema is indeed the most privacy-respecting app) would require separate ground-truth-based studies for each of the aforementioned ranking criteria, which falls outside the scope of this work.
\section{Discussion}
\label{sec:discussion}

\subsection{General findings}
 
We synthesize general findings (GF) that connect the three RQs with the main goal of the study and the three use cases introduced in Section~\ref{sec:method}, explicitly linking to findings F1--F11. 
 
\begin{tcolorbox}[insightboxrq4]
\textbf{GF1. LLM-based ranking rationale is broader than ASO signals but lacks standardization for practical use.}
LLMs address diverse ranking criteria, often going beyond standard ASO metrics, but these rationales are fragmented, inconsistently reported, and subjectively interpreted by different LLMs.
\end{tcolorbox}
 
The elicitation results from RQ\textsubscript{1} show that LLMs reference a wide range of criteria, many of which are not captured by conventional ASO practices (F2), often incorporating qualitative or user-centric indicators (F3). This breadth and lack of standardization is the starting point for the stakeholder-specific implications discussed in GF2--GF4 below. 
At the taxonomy level, the convergence of the consolidated criteria across cohorts and datasets (F1) suggests that standardization is feasible: the conceptual core of LLM ranking rationale appears shared, and only its surface expression is fragmented. 
This reading, however, follows from convergence in \emph{reported} criteria only; confirming it would require directly comparing LLM recommendations with those surfaced by app stores for the same queries, which our design does not include and which we flag as an assumption for future work rather than an established conclusion. To address the fragmentation we do observe, LLM-based recommenders should adopt structured formats for expressing ranking rationales through controlled vocabularies and slot-based templates, which could serve as anchors for interpreting recommendations, benchmarking model behaviour, and aligning developer efforts with observable LLM outputs.
 
\begin{tcolorbox}[insightboxrq4]
\textbf{GF2. Trust in LLM app recommendations must be calibrated jointly on rationale breadth, output stability, and criterion conditioning.} However can be assumed favourable by default.
\end{tcolorbox}
 
For end-users (UC1), the three RQs jointly determine how much trust a given recommendation deserves. 
The broader, more subjective ranking rationale identified in RQ\textsubscript{1} (F3) can capture personal, context-dependent preferences better than fixed ASO signals, but only if the user can identify which criteria a model is actually applying, which our results show is not guaranteed (GF1). 
Stability, in turn, is not a fixed property of ``the LLM'': it depends jointly on which model and cohort is queried (F5, F6) and which app domain is targeted (F7), so the same user request can yield highly reproducible answers in one context (e.g., mainstream streaming or messaging apps) and near-arbitrary ones in another (e.g., niche utilities). 
Finally, RQ\textsubscript{3} shows that asking a model to justify its ranking with an explicit criterion does change the recommendations it returns (F9), but this should not be read as increased reliability: guided recommendations are, on average, less reproducible than unconditioned ones (F10), and we cannot confirm from our data that the displaced recommendations genuinely satisfy the requested criterion. Users seeking transparency should therefore treat model self-reported rationale, including criterion-conditioned answers, as indicative rather than verified.
 
\begin{tcolorbox}[insightboxrq4]
\textbf{GF3. Broad market visibility remains the most reliable lever for LLM-driven discoverability.} Positioning around specific qualitative criteria is currently unverifiable and provider-specific.
\end{tcolorbox}
 
For developers and ASO specialists (UC2), RQ\textsubscript{1} shows that LLMs reward criteria that traditional ASO metadata does not directly encode (F2), e.g., interface quality, integration, or engagement, which in principle expands the surface on which an app could be positioned. However, RQ\textsubscript{3} shows that the criteria under which models actually agree with one another are those that restate popularity and engagement (F11); criteria oriented towards more specific, verifiable qualities (e.g., security, pricing) instead produce divergent, model-specific rankings (F10, F11). This implies that, at present, broad market visibility remains the most dependable route to presence in LLM-generated lists, while strategies targeting a specific qualitative criterion cannot be assumed to generalize across providers. We flag this as an assumption for developers to validate empirically for their own app and criterion of interest, since our design does not evaluate whether displaced recommendations are objectively aligned with the requested criterion. Additionally, since web search does not materially change either the elicited criteria (F4) or recommendation consistency (F8) for the proprietary cohort, developers should not expect that optimizing live web presence alone will shape LLM recommendation behaviour, at least at the level of ranking logic (see GF5).
 
\begin{tcolorbox}[insightboxrq4]
\textbf{GF4. Consistent, controllable, and verifiable LLM-based recommendation cannot be achieved through prompting alone.} It requires mechanisms external to the model.
\end{tcolorbox}
 
For recommender-system researchers (UC3), RQ\textsubscript{2} establishes that recommendation consistency is bounded jointly by model openness and app-domain structure (F5--F7), and RQ\textsubscript{3} shows that explicit ranking criteria, the most direct prompt-level lever available, do not lift this bound: conditioning reduces rather than increases both within-model determinism and cross-model agreement (F10), and its effect on displaced recommendations cannot be verified as criterion-aligned from outputs alone (F9). 
Together, this suggests that designing transparent and controllable app recommenders requires mechanisms external to prompting, such as aggregation across runs, cross-model ensembling, or retrieval-based grounding of each criterion in verifiable data. 
We put these forward as candidate directions rather than as validated solutions, since none is evaluated in this study. 
Model- and criterion-specific behavioural differences (e.g., \texttt{mistral-large-2512} and \texttt{claude-opus-4-6} combine high internal consistency with low sensitivity to conditioning, while \texttt{gemini-3-flash} and \texttt{llama4:scout} combine lower stability with aggressive restructuring) further suggest that stable, high-agreement recommendations and criterion-driven, exploratory ones could be combined within a single system through agreement-aware strategies, e.g., restricting recommendations to high-consensus items in trust-sensitive scenarios, and relaxing agreement thresholds for discovery-oriented ones. 

We note that these behavioural profiles are observed only for the models and settings studied here and may shift as models are updated, so they should not be read as fixed model categories. Both our evaluation methodology and the resulting consistency baselines can be directly reused to assess whether future contributions (e.g., fine-tuning, structured post-processing, retrieval grounding) genuinely improve controllability, rather than merely shifting outputs.
 
\begin{tcolorbox}[insightboxrq4]
\textbf{GF5. The observed recommendation dynamics are dominated by parametric knowledge; web augmentation adds cost and variability without changing them.}
Across both objects of analysis (i.e., ranking rationale and recommendation consistency) enabling web search does not materially alter model behaviour.
\end{tcolorbox}
 
The two information access ablations converge: web search neither expands the elicited ranking criteria (F4) nor changes the relative stability ranking of models, introducing instead a small, model-dependent increase in variability (F8). This indicates that the dynamics characterized in this study are properties of parametric knowledge and alignment rather than of retrieval pipelines, at least at the level of ranking rationale and consistency (see GF3 for the implication for developers, UC2, and GF4 for the resource-efficiency implication for researchers, UC3). We emphasize that this conclusion is scoped to criteria and consistency: web retrieval may well affect which specific apps are recommended and their factual currency, which our study deliberately does not assess.

\subsection{Threats to Validity}
\label{sec:ttv}

We organize threats to validity following the taxonomy by Wohlin et al.~\citep{Wohlin2012}.

\textit{Construct validity} threats stem from our feature datasets, taxonomy, and evaluation metric. 
Our datasets (16 features from 8 categories~\citep{Dabrowski2023} and 100 features from 18 categories~\citep{Jiang2019}) may not cover all recommendation scenarios, though drawing on two independent sources allows triangulation. 
Manual filtering of ranking criteria involved researcher judgement, and RQ\textsubscript{3} experiments use only the small-scale dataset for tractability; broader replication would strengthen convergence conclusions. 
Our taxonomy of 16 ranking criteria, consolidated from 76,119 raw outputs, favours generalization over exhaustivity, and may underrepresent niche reasoning patterns.
RBO assumes top-weighted rank overlap reflects consistency. However, rankings were stable across $p \in \{0.8, 0.9, 1.0\}$, and a name-normalization check changed the mean delta only marginally ($-0.038$ to $-0.034$), indicating naming variability does not drive our conclusions.

Concerning \textit{internal validity}, our ten LLMs differ in architecture, scale, and training data, which may affect comparability. 
The open-source cohort's 24~GB VRAM constraint partially confounds openness with scale, so the cohort stability gap (F5) may partly reflect smaller model size. 
We standardized prompts, schemas, and configurations, kept default temperatures to reflect typical usage, and note that results are a temporal snapshot subject to provider-side updates, motivating the longitudinal replications enabled by our package. 
Embedding-based clustering of criteria is sensitive to phrasing; we mitigated this with four validation metrics (silhouette analysis~\citep{Dinh2019}, gap statistic~\citep{Tibshirani2002}, elbow method~\citep{Shi2021}, Calinski-Harabasz~\citep{Caliński01011974}) plus manual inspection. 
Furthermore, there is no guarantee that LLM-stated criteria reflect actual selection drivers, consistent with documented unfaithfulness in LLM-generated explanations~\citep{Turpin2023,Lanham2023}. We treat this as a scope decision: F1–F11 characterize reported rationale, which is the only decision signal visible to users and developers in practice, but it does not reflect ground-truth internal reasoning.

\textit{External validity} is limited by our model and domain sampling: findings may not generalize to larger open-source models, fine-tuned or agentic pipelines, or retrieval setups beyond provider-native web search. Our scope is limited to English, single-turn, feature-based search over popular app domains, excluding multi-turn workflows, niche categories, or non-English ecosystems. Our methods and datasets are designed for replication and extension, including to domains such as software development tool and library recommendation~\citep{Sun2020}.


Finally, concerning \textit{conclusion validity}, all comparisons average across repeated runs (10 per feature for blind experiments, 5 per feature--criterion for guided experiments), balancing variability capture against the cost of large-scale multi-model querying. For RQ$_3$, we complement descriptive statistics with two-sided Wilcoxon signed-rank tests, Holm-corrected across criteria, with effect sizes; the 16-feature pairing limits statistical power, so non-significant moderate reductions should be interpreted as inconclusive rather than as evidence of no effect. 
Given the 16-feature pairing, we treat significance testing as confirmatory: effect sizes (Table~\ref{tab:rq3-per-criterion}) and the full RBO distributions (Figures~\ref{fig:rq3-displacement}, \ref{fig:rq3-external}) are the primary evidence for F9--F11, with significance testing used only to flag the subset of criteria where the effect is large enough to survive correction even at this sample size.
\section{Related Work}
\label{sec:related-work}

Several studies have explored the behaviour, potential and performance of LLMs as recommender systems. Due to its increased interest, Zhao et al. conducted a survey on the use and adaptation of LLMs for multiple recommendation tasks~\citep{Zhao2024}, including top-k recommendation~\citep{Sharma2024}, rating prediction~\citep{Kim2025}, conversational recommendation~\citep{Liu2023a}, and explanation generation~\citep{Lubos2024}. Complementarily, Xu et al. proposed a framework for adapting LLMs as recommender systems~\citep{Xu2025}, identifying four common paradigms: prompting without tuning~\citep{Liu2023}, full-model fine-tuning~\citep{Lin2024}, parameter-efficient fine-tuning~\citep{Bao2023}, and instruction tuning~\citep{Luo2025}. Our study focuses on the first (i.e., prompting without tuning), replicating the common scenario for most users accessing commercial LLMs. More recently, Peng et al. conducted a survey on agentic recommender systems~\citep{Peng2025}, focusing on methods for recommendation, agent components, datasets and metrics for evaluation. These contributions align design decisions, development frameworks, and evaluation criteria beyond accuracy for multiple recommendation tasks. However, existing surveys and evaluation studies largely focus on recommendation effectiveness or adaptation strategies, leaving comparatively little understanding of how recommendation behaviour varies across proprietary and open-source LLMs and across knowledge-only versus web-augmented settings.

In alignment with our work, empirical studies have explored the behaviour of vanilla conversational LLMs as recommender systems. Hou et al. investigated the capacity of ChatGPT's \texttt{gpt-3.5-turbo} model to act as a zero-shot recommender system~\citep{Hou2024a}. They focused on exploring factors affecting ranking performance, as well as the data and knowledge on which such models rely. While they demonstrated the potential of LLMs as zero-shot recommenders, they also reported that the popularity of ranking items (e.g., mobile apps) can bias recommendation results independently of prompt details, while arguing that popularity itself is difficult to measure. This reinforces the relevance of examining how external information access may influence LLM recommendation behaviour and reported ranking rationales, which we investigate through a controlled comparison of knowledge-only and web-augmented settings. Dai et al. conducted a comprehensive evaluation of ChatGPT's recommendation performance across different ranking paradigms~\citep{Dai2023}, showing that ranking behaviour is sensitive to prompt formulation and ranking paradigms across domains. Their insights highlight the importance of carefully controlling prompt design and recommendation depth, both of which we incorporate into our experimental pipeline for eliciting mobile app recommendations.

More recently, Manzoor et al. conducted an online user study with 190 participants assessing the performance of ChatGPT as a movie recommender system~\citep{Manzoor2024}. They focused on multiple perceived quality descriptors, including recommendation accuracy, explainability, adequacy, and language quality. In particular, ChatGPT showed the highest quality improvements in explainability and language quality compared to a baseline contextual retrieval-based method. Similarly, Lubos et al. explored the ability of LLMs to generate personalized, context-sensitive explanations for software recommendations, confirming their utility beyond item ranking tasks~\citep{Lubos2024}. These studies support our decision to focus on zero-shot prompting and domain-sensitive instructions as viable strategies for analysing LLM recommendation behaviour without additional training.

In the field of LLM-based software recommendation, Ma et al. proposed CAPIR~\citep{Ma2024}, a compositional framework that decomposes coarse-grained requirements into subtasks and retrieves relevant APIs using prompting alone. Their approach also relies on zero-shot prompting and document-based retrieval, excluding fine-tuning. John et al. introduced LLMRS~\citep{John2024}, which ranks software products using sentiment scores from user reviews. While reliant on quantitative aggregation (i.e., computing positivity scores across reviews), their work similarly confirms the viability of zero-shot LLM-based recommendation. 
In the mobile app domain, Maji et al. released MobileConvRec~\citep{Maji2024}, a dataset combining user interaction sequences with simulated multi-turn dialogues. While their goal is training dialogue-based recommenders, our focus is on assessing pre-trained LLMs in realistic usage scenarios. Lastly, Maqbool et al. proposed INTERPOS~\citep{Maqbool2025}, which enhances recommendation accuracy by modelling user interaction rhythms (i.e., time intervals between consecutive user actions). While these studies target various aspects of LLM-based recommendation, our work provides a systematic observational analysis of prompt-driven mobile app recommendations, focusing on the ranking criteria reported by LLMs, the consistency of their recommendations, and the effects of model openness and information access on recommendation behaviour.
\section{Conclusions}
\label{sec:conclusions}

This study examined how general-purpose LLMs generate mobile app recommendations through prompt-based interactions, covering ten models across proprietary and open-source cohorts, two feature datasets of different scale, and both web-augmented and knowledge-only information access settings. 
We presented a taxonomy of ranking criteria elicited from LLM outputs, a systematic evaluation of internal and external recommendation consistency, and an assessment of whether conditioning on explicit ranking criteria steers models towards convergent recommendations. 
Our results show that LLMs apply a wide range of criteria, often different and possibly richer than established ASO metrics, though sometimes inconsistently. 
These findings suggest that while LLMs can act as flexible, general-purpose recommenders and surface novel considerations beyond conventional ranking signals, their outputs require careful interpretation --- especially when used in decision-support scenarios. 
Hence, our study highlights the need for increased transparency and more systematic evaluations of recommendation logic.

Building on these findings, our study offers several novel insights that enrich the understanding of LLM-based recommendation. 
First, we reveal that general-purpose LLMs spontaneously construct ranking rationales that are broader and more human-centric than standard ASO signals, capturing qualities such as \textit{Security and Privacy}, \textit{User Interface and Experience}, or \textit{Customer Support} without explicit training for this domain. 
This rationale converges across cohorts and datasets, and web search access does not materially modify it. 
Second, we show that recommendations can be strategically steered through criteria-conditioned prompting, enabling guided discovery of apps that would remain hidden in blind recommendations. 
However, steering does not entail control: conditioning reduces both within-model determinism and cross-model agreement, indicating that the agreement observed among models rests on a shared popularity-driven default rather than on shared criterion-specific knowledge. 
Third, by mapping stability patterns across runs, models, and domains, we expose how recommendation consistency is jointly bounded by model openness and app-domain structure, a coexistence of stability and diversity that can be harnessed to balance trust and exploration. 
These contributions show that LLMs can articulate nuanced, multi-dimensional perspectives on software quality, extending well beyond the capabilities of traditional recommender pipelines. 
At the same time, our findings suggest that LLM-based recommendations should not be interpreted as replacements for traditional ASO-driven ranking systems, but rather as complementary mechanisms that expose user-centric criteria not explicitly captured by conventional ranking signals. 
Combining these perspectives can help stakeholders reconcile quantitative ranking indicators with the qualitative assessments produced by LLMs.

Future work should explore how LLM-based recommendations can be made more robust and interpretable, particularly by aligning stated ranking criteria with measurable ground truth data and by grounding criteria-conditioned prompting in verifiable, retrieval-based evidence. 
Additional efforts could compare LLM-generated recommendations with app store rankings for the same queries, assess how users perceive the reliability of LLM-generated app rankings, and track the longitudinal evolution of recommendation behaviour across model versions. 
While this study focused on mobile apps, the proposed methodology and replication package are domain-agnostic, and can be extended to support empirical studies on LLM-based recommendation for other types of software artifacts, larger open-source models, and interactive multi-turn scenarios. 
The included use cases demonstrate how our approach supports practical and reproducible assessment across multiple recommendation scenarios.
\section*{Acknowledgments}

\section*{Data Availability Statement}

Datasets, source code, and evaluation results are available at \url{https://github.com/nlp4se/llm-app-recommender}.

\section*{Declaration of generative AI and AI-assisted technologies in the manuscript preparation process}
\label{sec:generative-ai}


During the preparation of this work, the authors used Claude Opus 4.6--5 models for language editing, clarity improvements, limited code assistance (e.g., code refactoring, debugging support, and implementation suggestions), and drafting documentation for the replication package. The authors reviewed and edited the output as needed and take full responsibility for the content of the published article. All research design, experimental methodology, data analysis, code validation, and interpretations remain the sole work and responsibility of the authors.


\bibliographystyle{cas-model2-names}
\bibliography{refs}

\end{document}